\documentclass[preprint,12pt]{elsarticle}
\usepackage{dblfloatfix}
\usepackage{amssymb}
\usepackage{amsmath}
\usepackage{dsfont}
\usepackage{bm}
\usepackage{placeins}

\begin{document}

\begin{frontmatter}
\title{Plasma Waves as a Benchmark Problem}

\author[a]{Patrick Kilian\corref{pkilian}}
\author[b]{Patricio A. Mu\~noz}
\author[a,c]{Cedric Schreiner}
\author[a]{Felix Spanier}

\cortext[pkilian]{Corresponding author.\\\textit{E-mail address:} 28233530@nwu.ac.za}

\address[a]{Centre for Space Research, North-West University, Potchefstroom, South Africa}
\address[b]{Max-Planck-Institute for Solar System Research, G\"ottingen, Germany}
\address[c]{Lehrstuhl f\"ur Astronomie, Universit\"at W\"urzburg, Germany}

\begin{abstract}
A large number of wave modes exist in a magnetized plasma. Their properties are
determined by the interaction of particles and waves. In a simulation code, the
correct treatment of field quantities and particle behavior is essential to correctly
reproduce the wave properties. Consequently, plasma waves provide test
problems that cover a large fraction of the simulation code.

The large number of possible wave modes and the freedom to choose parameters
make the selection of test problems time consuming and comparison between different
codes difficult. This paper therefore aims to provide a selection of test
problems, based on different wave modes and with well defined parameter values,
that is accessible to a large number of simulation codes to allow for easy
benchmarking and cross validation.

Example results are provided for a number of plasma models. For all plasma
models and wave modes that are used in the test problems, a mathematical
description is provided to clarify notation and avoid possible
misunderstanding in naming.
\end{abstract}


\end{frontmatter}

\section{Introduction}

Testing simulation codes for correct implementation and a sufficiently accurate
representation of reality is an important task. Our purpose is to make this task easier for
codes that simulate collisionless plasmas and hopefully lead to more widespread
validation activity. To this end, we propose a test problem that can be used for benchmarking
of different simulation codes, as well as validation with respect to analytic
solutions of the PDEs describing the system. This paper describes the setup,
analysis and parameters in all the details that are necessary for an independent
replication and comparison with other codes. To make the test
as accessible and useful as possible, we choose parameters that allow for
simulations with many different methods while minimizing the computational effort.
To the latter end, the proposed test does not rely on more than one spatially resolved
dimension. While invariance under exchange of axis can be tested (and to some
degree isotropy of wave propagation), it is expedient to complement this set of test
problems with other tests that directly check for effects of two or three
spatial dimensions.

The test case proposed here uses plasma wave modes in a homogeneous plasma. No
complicated setup or boundary conditions are necessary and the corresponding
theory is well established and widely available. From a numerical side, however,
wave modes are an interesting test as the properties of the wave are determined
by the interaction of the particles in the plasma with the electromagnetic
fields. Consequently, a large part of the simulation code is covered by this integration test, verifying
both the internal consistency of the code and the correct interaction of
the different modules. Some classes of problems in the code lead to
characteristic deviations in the simulation results. With sufficient experience, they can be tracked back to
the responsible part of the code, but the process
can be difficult and time consuming. The test is therefore not meant to replace
unit tests that check individual functions, but to complement them. The design of
the test is such that the numerical effort is low enough that it can be
run before any checking into a source control system to guard against
regressions.

This paper is of course not the first test case that is available to the
simulation community. Probably the most widely used benchmark for comparisons
between plasma simulation codes is the GEM reconnection challenge
\cite{Birn_2001}. This setup has been simulated by Magneto-Hydro-Dynamic (MHD),
Hall MHD, hybrid codes using kinetic ions and an electron fluid with or without
inertia, and Particle-in-Cell (PiC) codes. Comparison between different codes has led
to a better understanding of the relevance of physical effects that are
represented to various degrees in different simulation methods. Beyond that, it is
a standard test case that is often used to measure the performance of new codes.
For the simulation of fusion devices, there is benchmarking and comparison
efforts such as \cite{Dimits_2000,Falchetto_2008,Goerler_2016} that simulate
properties of fusion plasmas such as the turbulence that is driven by the
thermal gradient between the core and the edge of these plasmas. The goal is to
improve the reliability of predictions based on simulations through the
cross-comparison of simulation codes.



\section{Plasma Model}

To validate a numerical code it is necessary to have a well specified
mathematical model of the system that the code is supposed to simulate. The
basic equations of each model are given here to clarify the nomenclature used
for the following discussion and to avoid any possible confusion about the
model and to avoid conflicts of notation. A more detailed explanation of the
physical meaning and implication of the equations can be found in standard
textbooks such as \cite{Jackson_1999} and \cite{Bittencourt_2004}. The
equations are given in terms of the (particle) velocity $\vec{v}$ instead of
the momentum $\vec{p}$, which is only appropriate in the non-relativistic
limit. For plasma temperatures and wave intensities discussed here this is,
however, sufficient.\footnote{The plasma temperature will be defined in the
description of the individual test problems. The wave intensities are set by
the intrinsic thermal and numerical noise and are much below the regime where
wave modes couple or show non-linear effects.}

For a collisionless plasma, the mathematical model is given by the Vlasov
equation \cite{Vlasov_1938}. This is an evolution equation for the phase space
density $f_\alpha$ of one particle species $\alpha$ as a function of position $\vec{r}$ and
velocity $\vec{v}$:
\begin{equation}
	\frac{\partial f_\alpha(\vec{r}, \vec{v}, t)}{\partial t} + \vec{v} \cdot \nabla_\mathrm{r} f_\alpha + \frac{1}{m_\alpha} \vec{F}(\vec{r}, \vec{v}, t) \cdot \nabla_\mathrm{v} f_\alpha = 0\quad ,
\label{eqn:vlasov}
\end{equation}
where $\vec{F}$ is the total force acting on $f$.

From the distribution function, we can also compute the source terms for the
electromagnetic fields by integrating over the velocity components of the phase
space. This leads to the (net) charge density $\rho$ and current density
$\vec{\jmath}$:

\begin{align}
	\rho(\vec{r}, t) &= \sum_\alpha q_\alpha \int f_\alpha(\vec{r}, \vec{v}, t) \;\mathrm{d}\vec{v}\quad,
	\\
	\vec{\jmath}\,(\vec{r}, t) &= \sum_\alpha q_\alpha \int \vec{v} \; f_\alpha(\vec{r}, \vec{v}, t) \;\mathrm{d}\vec{v}\quad.
\label{eqn:source_terms}
\end{align}

The reaction of the particles with charge $q_\alpha$ to the fields is given by
the Lorentz force (\cite{Lorentz_1892}). In Gaussian cgs units the electric
field $\vec{E}$ and the magnetic field $\vec{B}$ exert the force

\begin{equation}
	\vec{F}(\vec{r}, \vec{v}, t) = q_\alpha \left(\vec{E}(\vec{r}, t)+ \frac{\vec{v}}{\mathrm{c}} \times \vec{B}(\vec{r}, t)\right)\quad.
	\label{eqn:lorentz}
\end{equation}

To close the set of equations, we need the evolution equations for the fields.
These are given by Maxwell's equations or some approximation thereof, depending
on the plasma model. They are discussed in the following subsections.

\subsection{Electromagnetic}
\label{subsec:electromagnetic}

The electromagnetic plasma model uses the full set of Maxwell's equations
\cite{Maxwell_1865}:
\begin{align}
	\nabla \times \vec{E}(\vec{r}, t) &= - \frac{1}{\mathrm{c}} \frac{\partial}{\partial t}\vec{B}(\vec{r}, t)\quad,
	\label{eqn:maxwell_gaussian}
	\\
	\nabla \times \vec{H}\!(\vec{r}, t) &= \phantom{-} \frac{1}{\mathrm{c}} \frac{\partial}{\partial t} \vec{D}(\vec{r}, t) + \frac{4 \pi}{\mathrm{c}} \vec{\jmath}\,(\vec{r}, t)\quad,
	\label{eqn:maxwell_gaussian_second}
	\\
	\phantom{\frac{\partial}{\partial t}}\nabla \cdot \vec{D}(\vec{r}, t) \; &= \phantom{-} 4 \pi \rho(\vec{r}, t)\quad,
	\\
	\phantom{\frac{\partial}{\partial t}}\nabla \cdot \vec{B}(\vec{r}, t) \; &= \phantom{-} 0\quad.
	\label{eqn:maxwell_gaussian_last}
\end{align}
Eqs.~\ref{eqn:maxwell_gaussian} to \ref{eqn:maxwell_gaussian_last} formally
involve the electric displacement $\vec{D}$ and magnetic intensity $\vec{H}$.
In vacuum -- without material effects -- these can be replaced by the electric
field $\vec{E}$ and magnetic induction $\vec{B}$ as the permittivity and
permeability of free space $\epsilon_0$ and $\mu_0$ are set to unity in the
units used. Without source terms, these equations have wavelike solutions that
describe electromagnetic radiation. In a plasma this is also the case, but the
wave is modified due to the interaction between particles and fields. Whenever
the interaction of electromagnetic radiation (e.g. radio waves or laser pulses)
with a plasma is of interest, the full Vlasov-Maxwell-system is the model of
choice. However, the high propagation speed of these waves lead to a very
restrictive limit on the permissible time step in explicit codes (see e.g.
\cite{Birdsall_2005}). Therefore it is often convenient to couple the Vlasov
equation to approximations of Maxwell's equations that do not allow for the
existence of light waves.

\subsection{Radiation-free}
\label{subsec:darwin}

There are several different ways to derive the radiation-free approximation to
Maxwell's equations. It can be seen as the correct approximation up to order
$\mathcal{O}\left(v^2 / \mathrm{c}^2\right)$. Alternatively, one can approach it
through a Helmholtz decomposition of the electric field and the current, split
into a longitudinal curl-free part and a transverse divergence-free part. The
displacement current, given by time derivative of the transverse electric
field, is removed from Amp\`ere's Law which removes light waves.
Ref.~\cite{Krause_2007} showed that either way leads to the set of equations:

\begin{align}
	\nabla \times \vec{B} & = \frac{1}{\mathrm{c}} \frac{\partial \vec{E}_\mathrm{L}}{\partial t} + \frac{4 \pi}{\mathrm{c}} \vec{\jmath}, \label{eqn:maxwell_darwin1} \\
\nabla^2 \vec{E}_\mathrm{T} & = \frac{4 \pi}{\mathrm{c}^2} \frac{\partial \vec{\jmath}_\mathrm{T}}{\partial t}, \label{eqn:maxwell_darwin2} \\
	\nabla \;\vec{E}_\mathrm{L} \, & = 4 \pi \, \rho \quad . \label{eqn:maxwell_darwin3}
\end{align}

This approximation to Maxwell's equations is also known as Darwin approximation
\cite{Darwin_1920} or as magnetoinductive model as the production of magnetic
fields from currents is retained. Only the effect of the transverse
displacement current is removed.

\subsection{Electrostatic}
\label{subsec:electrostatic}

For particle velocities much slower than the speed of light and without large
scale currents, one can completely remove the transverse electric field. This
way, one obtains the electrostatic model, where the magnetic field is constant in
time and the longitudinal electric field is given by
Eq.~\eqref{eqn:maxwell_darwin3}. For this plasma model, no current has to be
calculated from the particle motion, as the charge density in each time step is
sufficient to calculate the fields. The downside is of course that wave modes
that require transverse fields are not present in this model.

\subsection{Implicit Electron Fluid}
\label{subsec:hybrid}

If kinetic effects of electrons are not important, it is possible to reduce the
numerical effort by treating the electrons as a fluid. The electron momentum
equation leads to a generalized Ohm's law for the electric field:
\begin{equation}
	\vec{E} = - \frac{m_\mathrm{e}}{e} \left( \frac{\partial \vec{u}_\mathrm{e}}{\partial t} + \left(\vec{u}_\mathrm{e} \cdot \nabla \right) \vec{u}_\mathrm{e} \right) - \frac{\vec{u}_\mathrm{e}}{\mathrm{c}} \times \vec{B} - \frac{\nabla \, P_\mathrm{e}}{e \, n_\mathrm{e}} + \frac{m_\mathrm{e}}{e} \, \nu \vec{\jmath}\quad .
	\label{eqn:emhd_electric_field}
\end{equation}
In this equation $\vec{u}_\mathrm{e}$ gives the flow speed of the electron
fluid, $P_\mathrm{e}$ its pressure and $\nu$ the resistivity.

Combining this equation with Faraday's law leads to an equation for the
magnetic field, or rather for the generalized vorticity $\vec{W}$:

\begin{align}
	\frac{\partial}{\partial t} \vec{W} &= \nabla \times \left(\vec{u}_\mathrm{e} \times \vec{W}\right) - \nabla \times \frac{\nabla P_\mathrm{e}}{m_\mathrm{e}\, n_\mathrm{e}} + \nu \, \nabla \times \vec{\jmath} \label{eqn:vorticityupdate}\quad , \\
	\vec{W} &= \nabla \times \vec{u}_\mathrm{e} - \frac{e}{m_\mathrm{e} \, c} \vec{B} \label{eqn:vorticitydef}\quad .
\end{align}

There are two regimes where this description is worth consideration. One is on
electron scales, where ions can be considered immobile due to their large
inertia and do not affect the dynamics of the system. This model is called
electron magnetohydrodynamics (EMHD) and is not considered in this paper.

Here, we are interested in the ions and the behavior of the system on their
timescales. Then we can assume that the electrons are sufficiently mobile to
neutralize their charge density nearly instantaneously. In this hybrid model of
kinetic ions and fluid electrons we use $n_e = n_i$ at any instance. The total
current that is determined by the magnetic field must then be carried either by
the ions or the bulk flow of the electron fluid:

\begin{equation}
	\frac{\mathrm{c}}{4 \pi} \nabla \times \vec{B} = \vec{\jmath}_\mathrm{i} + \vec{\jmath}_\mathrm{e} = \vec{\jmath}_\mathrm{i} - e \, n_\mathrm{e} \, \vec{u}_\mathrm{e} \label{eqn:currentbalance} \quad .
\end{equation}

The current carried by the ions can be determined based on the marker particles
and deposited onto the grid. Inserting Eq.~\eqref{eqn:currentbalance} into
Eq.~\eqref{eqn:vorticitydef} removes the unknown flow speed of electrons and
leads to an equation that can be solved for $\vec{B}$
numerically.\footnote{Once the magnetic field is available,
$\vec{u}_\mathrm{e}$ can of course be calculated as well and added to the
output, to study the motion of the electron fluid.}

The hybrid model described so far contains effects due to finite electron
inertia. Most hybrid codes ignore this effect as the ion mass is much larger
than the electron mass. The hybrid code here also allows to do so and the
reference results for the test problems are provided with and without electron
inertia.

\section{Small Amplitude Waves}
\label{sec:small_amplitude_waves}

To find the well known small amplitude waves \footnote{These solutions to the
linearized wave equation are also known as linear modes or eigen modes.} of a
plasma model, it is necessary to linearize its governing equations. That means
we assume that all quantities can be split into a static and homogeneous
background part (subscript $0$) and a small perturbation (subscript $1$). For
physical reasons, we assume that $\vec{E}_0 = 0$ and $\vec{\jmath}_0 = 0$ and drop any
term that contains two or more factors with subscript one, because we expect
such contributions that are second order in the small perturbation to be
negligible. Furthermore we can, without loss of generality, align our
coordinate system in such a way that the static and homogeneous background
magnetic field $\vec{B}_0$ is aligned with the $z$ axis and the propagation
direction of the wave $\vec{k}$ is in the $x$-$z$-plane.

Even with those simplifications, it is hard to self-consistently solve the Vlasov-Maxwell-system.
The canonical method by Landau \cite{Landau_1946} requires
Laplace transformations and residue calculus for integration in the complex
plane. Details can be found e.g. in \cite{Koskinen_2011}.

For illustration, it suffices to analyze the electromagnetic case, neglecting
any thermal effects and assuming that the perturbations are plane waves that can be
written as (possibly complex) constants times $\exp\left(\imath(\vec{k} \cdot
\vec{r} - \omega\, t)\right)$. For this harmonic case, the first two Maxwell's
equations (Eqs.~\ref{eqn:maxwell_gaussian}-\ref{eqn:maxwell_gaussian_second})
reduce to:

\begin{align}
	\vec{k} \times \vec{E}(\vec{r}, t) &= \phantom{-} \frac{\omega}{\mathrm{c}} \vec{B}(\vec{r}, t)
\quad,
	\label{eqn:harmonic_curlE}
	\\
	\vec{k} \times \vec{B}(\vec{r}, t) &= - \frac{\omega}{\mathrm{c}} \vec{E}(\vec{r}, t) - \frac{4 \pi \, \imath}{\mathrm{c}} \vec{\jmath}\,(\vec{r}, t)\quad.
	\label{eqn:harmonic_curlB}
\end{align}

Assuming a linear response of the plasma to the electric field, we can write the
current density $\vec{\jmath}$ as $\bm{\sigma} \, \vec{E}$ using the conductivity tensor
$\bm{\sigma}$. Inserting this into Amp\~ere's law given by Eq.~\eqref{eqn:harmonic_curlB}, leads to
\begin{equation}
	\vec{k} \times \vec{B} = - \frac{\omega}{\mathrm{c}} \, \bm{\epsilon} \, \vec{E}\quad,
\end{equation}
using the dielectric permittivity tensor $\bm{\epsilon}$:
\begin{equation}
	\bm{\epsilon} = \mathds{1} + \frac{4 \pi \,i \, \bm{\sigma}}{\omega}\quad.
\end{equation}

Using Faraday's law given in Eq.~\eqref{eqn:harmonic_curlE},
we can substitute for the magnetic field and obtain:

\begin{equation}
	\vec{n} \times \vec{n} \times \vec{E} + \bm{\epsilon} \, \vec{E} = 0\quad.
	\label{eqn:curl_curl_E}
\end{equation}

The vector $\vec{n} = \mathrm{c} \, \vec{k} / \omega$ is a scaled version of
the wave vector $\vec{k}$, which is dimensionless and its magnitude corresponds to the
refractive index of the medium.

Rewriting Eq.~\eqref{eqn:curl_curl_E} once more we get
\begin{equation}
	\mathbf{D}(\omega, \vec{k}) \, \vec{E} = 0 \quad ,
	\label{eqn:dielectriceqn}
\end{equation}
where $\mathbf{D}$ is the dielectric tensor. For this equation to have a
nontrivial solution, the determinant of the 3x3 tensor has to vanish.

Before going further, it is useful to introduce three quantities for each
species present in the plasma. These are the plasma
frequencies $\omega_{p,\alpha}$, the gyro frequencies $\Omega_\mathrm{c,\alpha}$ and
the sign of the charge $s_\alpha$. They are given by

\begin{align}
	\omega_{\mathrm{p},\alpha} &= \sqrt{\frac{4 \pi \, n_\alpha \, q_\alpha^2}{m_\alpha}}\quad,
	\\
	\Omega_{\mathrm{c},\alpha} &= \frac{|q_\alpha| \, |\vec{B_0}|}{m_\alpha \, \mathrm{c}}\quad,
	\\
	s_\alpha &= \frac{q_\alpha}{|q_\alpha|}\quad.
\end{align}

It is useful to introduce the usual Stix parameters \cite{Stix_1962} before
discussing the different solutions of Eq.~\eqref{eqn:dielectriceqn}:

\begin{align}
	R &= 1 - \sum_\alpha \frac{\omega_{\mathrm{p},\alpha}^2}{\omega^2} \cdot \frac{\omega}{\omega + s_\alpha \Omega_{\mathrm{c},\alpha}}\quad,
	\\
	L &= 1 - \sum_\alpha \frac{\omega_{\mathrm{p},\alpha}^2}{\omega^2} \cdot \frac{\omega}{\omega - s_\alpha \Omega_{\mathrm{c},\alpha}}\quad,
	\\
	P &= 1 - \sum_\alpha \frac{\omega_{\mathrm{p},\alpha}^2}{\omega^2}\quad,
	\\
	S &= \frac{1}{2} \left(R + L\right)\quad,
	\\
	D &= \frac{1}{2} \left(R - L\right)\quad.
\end{align}

Using the Stix parameters and the angle $\vartheta$ between the background
magnetic field $\vec{B}_0$ and the wave normal vector $\vec{n}$, the dielectric
tensor in Eq.~\eqref{eqn:dielectriceqn} reads:

\begin{equation}
	\mathbf{D}(\omega, \vec{k}) = \left( \begin{array}{ccc} S - n^2 \cos^2 \vartheta & -\imath D & n^2 \cos\vartheta\sin\vartheta\\\imath D& S-n^2&0\\n^2\cos\vartheta\sin\vartheta&0&P-n^2\sin^2\vartheta\end{array}\right)\quad.
\label{eqn:explicitdielectrictensor}
\end{equation}

The dependence on $k$ is of course hidden in the index of refraction $n$
and all entries in the tensor $\mathbf{D}$ are frequency dependent. Each solution to
Eq.~\eqref{eqn:dielectriceqn} connects both $\omega$ and $\vec{k}$ in the form
of a dispersion relation that is characteristic for the wave mode.

The following test problems make use of a range of different wave modes. There
are two reasons why no single wave mode is sufficient. The first is that
different wave modes might use different parts of the simulation code.
Especially in the radiation free plasma model, longitudinal and transverse
fields are treated very differently and are best tested with two different wave
modes. The other reason is that no single wave mode makes for a good test for
every plasma model. To test an electromagnetic model, the electromagnetic mode
is computationally cheapest, but this mode is removed from all other plasma
models. On the other hand, low frequency modes such as ion Bernstein modes
require very expensive simulations in an explicit electromagnetic code that are
not practical.

Tab.~\ref{tab:test_matrix} at the end lists which wave modes are suitable for
each plasma model, allowing for a quick selection of the relevant description
following below.

\subsection{Electromagnetic Mode}
\label{subsec:wave:electromagnetic}

The first wave we want to consider is the electromagnetic wave. It does also
exist as a solution to Maxwell's equations in vacuum, where it has the trivial
dispersion relation

\begin{equation}
	\omega = \mathrm{c} \, k\quad.
	\label{eqn:disp_em_em}
\end{equation}

To get the equivalent dispersion relation in a plasma, let us first consider the
case without a background magnetic field. In that case, $D$ vanishes and $P = R
= L = S = 1- \omega_\mathrm{p}^2 / \omega^2$, with the joint plasma frequency
$\omega_\mathrm{p}$ of all species given by:

\begin{equation}
	\omega_\mathrm{p} = \sqrt{\sum_\alpha \omega_{\mathrm{p},\alpha}^2}\quad.
\end{equation}

This simplifies the Maxwell tensor significantly and the resulting
characteristic polynomial can be solved, yielding the following dispersion
relation for the electromagnetic wave:

\begin{equation}
	\omega^2 = \omega_\mathrm{p}^2 + \mathrm{c}^2 \, k^2\quad.
	\label{eqn:disp_em}
\end{equation}

Eq.~\eqref{eqn:disp_em} indicates that this wave has a low frequency cutoff at
the plasma frequency $\omega_\mathrm{p}$ and extends to arbitrarily large
frequencies. In numerical practice, there is a high frequency limit from the
Nyquist frequency that the grid imposes on the wavelength\footnote{Depending on
the numerical implementation, the frequency limit imposed by the finite time
steps might occur before that.}. The high frequency nature makes the
electromagnetic mode very suitable for a quick check of a simulation code implementing the electromagnetic plasma model, as
relatively few time steps are needed.

\subsection[Birefringence]{Magnetic Birefringence\footnote{This is the usual term in
\label{subsec:wave:hf_lr}
optics for the optical property of a medium that two waves of identical
frequency but different polarization experience different indices of
refraction.}}

The addition of a background magnetic field splits the electromagnetic mode
into two modes. For parallel propagation ($\vartheta = 0$) these are the L and
R mode.  Their respective dispersion relations are:

\begin{equation}
	n^2 = L, \hspace{1cm} n^2 = R\quad.
	\label{eqn:disp_em_lr}
\end{equation}

The L mode is left hand circularly polarized, while the R mode right handed. This
motivates their name and the name of the corresponding Stix parameter. Solving
for $\omega(k)$ is possible, but leads to rather long expressions. Both modes
behave very much like the electromagnetic mode but with a shifted cutoff. Their
cutoff frequencies are given by

\begin{align}
	\omega_\mathrm{cut,L} &= \frac{1}{2}\left(\left(\Omega_\mathrm{c,i} - \Omega_\mathrm{c,e}\right) + \sqrt{\left(\Omega_\mathrm{c,e}+\Omega_\mathrm{c,i}\right)^2 + 4 \;\omega_\mathrm{p}^2}\right)\quad,\label{eqn:cutoff_L}\\
	\omega_\mathrm{cut,R} &= \frac{1}{2}\left(\left(\Omega_\mathrm{c,e} - \Omega_\mathrm{c,i}\right) + \sqrt{\left(\Omega_\mathrm{c,e}+\Omega_\mathrm{c,i}\right)^2 + 4 \;\omega_\mathrm{p}^2}\right)\label{eqn:cutoff_R}\quad.
\end{align}

In the limit of $\omega_\mathrm{cut} \gg \Omega_\mathrm{c,e} \gg
\Omega_\mathrm{c,i}$, the right hand side of
Eqs.~\eqref{eqn:cutoff_L}-\eqref{eqn:cutoff_R} simplifies to $\omega_\mathrm{p} \pm
1/2 \;\Omega_\mathrm{c,e}$. The L and R modes have a second branch at much
lower frequencies, below the gyro frequencies of ions and electrons
respectively. These exist even in radiation free plasma models and provide a
suitable test problem for them.

In this range of frequencies, that is especially relevant for EMHD or hybrid simulations,
the dispersion relation of the right hand circular mode can be found in e.g.
\cite{Bulanov_1992} or \cite{Shaikh_2009} and is given by:

\begin{equation}
	\omega = \Omega_\mathrm{c,e} \, \frac{d_\mathrm{e}^2 k^2}{1 + d_\mathrm{e}^2 k^2} \quad .
\label{eqn:disp_r_emhd}
\end{equation}

The product of electron skin depth $d_\mathrm{e}$ and the wave number $k$ can
be expected to be not too large. In the limit $k\,d_\mathrm{e} \gg 1$ the wave
frequency has the limiting value $\omega \rightarrow \Omega_\mathrm{c,e}$,
however the wave is usually absorbed before that.

In a hybrid model without electron inertia the nature of the low frequency R mode
changes. To derive the dispersion relation, it is necessary to insert the
definitions of gyro frequency $\Omega_\mathrm{c,e}$ and electron skin depth
$d_\mathrm{e}$ into Eq. \eqref{eqn:disp_r_emhd} and take the limit
$m_\mathrm{e} \rightarrow 0$. Doing so leads to
\begin{equation}
	\omega = \frac{\mathrm{c}\,B}{4 \pi \, n_\mathrm{e} \, e} k^2 \quad ,
\label{eqn:disp_r_emhd_massless}
\end{equation}
which is well defined even in the absence of electron inertia. However, at
larger $k$ it lacks the cutoff at the electron gyro frequency, which is not
well defined without electron mass.

\subsection{Extraordinary Mode}
\label{subsec:wave:extraordinary}

In the case of perpendicular propagation (i.e. $\vartheta = \pi / 2$), we also
find that the electromagnetic mode is split into a pair of modes. The mode
where the electric field component is parallel to the background magnetic field
behaves just as in the unmagnetized case. This is called the ordinary mode (or
in short O mode). The other mode, which only exists in the presence of a static
magnetic field, is called extraordinary mode (or X mode). Its electric field
component is perpendicular to both $k$ and $B_0$. The dispersion relation of
this second mode is given by
\begin{equation}
	\frac{\mathrm{c}^2 k^2}{\omega^2} = \frac{\left(\left(\omega+\Omega_\mathrm{c,i}\right)\left(\omega-\Omega_\mathrm{c,e}\right)-\omega_\mathrm{p}^2\right)\left(\left(\omega-\Omega_\mathrm{c,i}\right)\left(\omega+\Omega_\mathrm{c,e}\right)-\omega_\mathrm{p}^2\right)}{\left(\omega^2 - \Omega_\mathrm{c,i}^2\right)\left(\omega^2-\Omega_\mathrm{c,e}^2\right)+\omega_\mathrm{p}^2\left(\Omega_\mathrm{c,e}\Omega_\mathrm{c,i} - \omega^2\right)}\quad.
	\label{eqn:disp_x_mode}
\end{equation}
This mode has a cutoff at
\begin{equation}
	\omega_\mathrm{cut,X} = \omega_\mathrm{cut,R} = \frac{1}{2} \left(\Omega_\mathrm{c,e} - \Omega_\mathrm{c,i} + \sqrt{\left(\Omega_\mathrm{c,e}+\Omega_\mathrm{c,i}\right)^2 +4 \omega_\mathrm{p}^2} \right) \quad .
\end{equation}
This frequency is close to the upper hybrid frequency which is given by
\begin{equation}
	\omega_\mathrm{UH} \approx \sqrt{\Omega_\mathrm{c,e}^2 + \omega_\mathrm{p}^2} \quad .
\end{equation}

Depending on the magnetization, a second branch of the extraordinary mode close
the plasma frequency exists. If it exists it has a lower cutoff frequency
\begin{equation}
	\omega'_\mathrm{cut,X} = \omega_\mathrm{cut,L} = \frac{1}{2} \left(\Omega_\mathrm{c,i} - \Omega_\mathrm{c,e} + \sqrt{\left(\Omega_\mathrm{c,e}+\Omega_\mathrm{c,i}\right)^2 +4 \omega_\mathrm{p}^2} \right) \quad .
\end{equation}

\subsection{Langmuir Mode}
\label{subsec:wave:langmuir}

If we return to the case without a magnetic background field and reexamine the
dielectric tensor as given in Eq.~\eqref{eqn:explicitdielectrictensor}, we
notice that the characteristic polynomial has three solutions. Two of them are
identical and belong to the electromagnetic mode -- discussed above --
with two independent degrees of freedom (either two linearly polarized modes or
equivalently two circularly polarized modes). However, there exists a third
solution to $\left|\mathbf{D}(\omega,\vec{k})\right| = 0$ which satisfies
$\omega = \omega_\mathrm{p}$. This describes plasma oscillations which -- for a cold
plasma -- have constant frequency, vanishing group velocity and are purely
electrostatic.

To get a wave mode with well defined propagation behavior and wavenumber
dependence, it is necessary to include the effects of a finite temperature in a
warm plasma. This can be done for any wave mode but is tedious as the
permittivity tensor needs to be redefined to include the distribution function.
In the case of a Maxwellian velocity distribution, this is generally expressed
using the plasma dispersion function $Z(\zeta)$ (see \cite{Fried_1961}). Solving the resulting equations
to get the dispersion relations is complicated by the extra terms or might even
be impossible to do analytically in the general case. For the test problems, it
is sufficient to proceed in a less rigorous manner with the Langmuir mode.
Assuming an electron temperature $T_\mathrm{e}$, we can modify the dielectric
permittivity tensor to include the leading term, resulting in:

\begin{equation}
	\epsilon = 1 - \frac{\omega_\mathrm{p}^2}{\omega^2} - 3 \, \frac{k^2 \, \omega_\mathrm{p}^2}{m_\mathrm{e} \, \omega^4} \, T_\mathrm{e} \quad .
\end{equation}

Solving for $\omega^2$ to get the dispersion relation, we end up with
\begin{equation}
	\omega^2 = \frac{1}{2} \left(\omega_\mathrm{p}^2 + \sqrt{\omega_\mathrm{p}^4 + 12 \, \frac{k^2 \, T_\mathrm{e}}{m_\mathrm{e} \, \omega_\mathrm{p}^2}}\right)\quad,
\end{equation}
which of course for infinitely small $k$ is just $\omega_\mathrm{p}^2$ and can be
approximated, for not too large $k$, by
\begin{equation}
	\omega^2 = \omega_\mathrm{p}^2 + 3 \, k^2 \, T_\mathrm{e} / m_\mathrm{e} \quad .
\end{equation}

At this point, it is useful to introduce the Debye length $\lambda_\mathrm{D}$. This is
the natural length scale below which the plasma might contain charge imbalances
and electrostatic fields. The (electron) Debye length is given by

\begin{equation}
	\lambda_\mathrm{D} = \sqrt{\frac{T_\mathrm{e}}{m_\mathrm{e} \, \omega_\mathrm{p}^2}} = \frac{v_\mathrm{th,e}}{\omega_\mathrm{p}} \quad .
\end{equation}

Using that we can rewrite the dispersion relation of the Langmuir mode as

\begin{equation}
	\omega^2 = \omega_\mathrm{p}^2 \cdot \left(1 + 3 \, k^2 \lambda_\mathrm{D}^2\right) \quad .
	\label{eqn:disp_langmuir}
\end{equation}

The second term in this formulation being the correction due to the finite
temperature and Debye length. This expression is reasonably accurate as long as
the second term is less than unity, which fortunately is the case as Langmuir
waves of higher $k$ are quickly damped by electron Landau damping (see
\cite{Dawson_1961}). This is a completely collisionless effect
resulting from the interaction of the Langmuir wave with kinetic electrons and
as such is not applicable when electrons are described as a fluid.

\subsection{Bernstein Modes}
\label{subsec:wave:electronbernstein}

For the final wave mode considered in this set of test problems, we again add a
background magnetic field and have a look at the longitudinal modes that
propagate perpendicular to the background field. These waves also exist only in
a plasma of finite temperature, because only there the gyro radius is finite. For
a particle of species $\alpha$ the gyro radius $r_\alpha$ is given by:

\begin{equation}
	r_\alpha = \sqrt{\frac{k_\mathrm{B} \, T_\alpha}{m_\alpha \, \Omega_\mathrm{c,\alpha}^2}} = \frac{v_\mathrm{th,\alpha}}{\Omega_\mathrm{c,\alpha}} \quad .
\label{eqn:thermal_gyro_radius}
\end{equation}

The original derivation of the dispersion relation can be found in
\cite{Bernstein_1958} and will not be repeated here. To write down the
dispersion relation it is useful to rescale the wavenumber using the gyro radius
as follows:

\begin{equation}
	\lambda_\alpha = k^2 r_\alpha^2\quad.
\end{equation}

This quantity is non-zero in a warm plasma, corresponding to the presence of
finite Larmor radius effects. If $\lambda_\alpha$ is small this mostly leads to
gyro resonances at integer multiples of the gyro frequencies. The situation
gets complicated if we cannot make this assumption. At least for the case of
electrostatic waves propagating exactly perpendicular to the background magnetic
field, we can find the dispersion relation e.g. in \cite{Swanson_2003}. Using
the different definition of thermal speed implicitly used in
Eq.~\eqref{eqn:thermal_gyro_radius} and rewriting in terms of quantities defined
in this paper, we get:

\begin{equation}
	1 - \sum_{\alpha} \frac{2\,\omega_\mathrm{p,\alpha}^2}{\lambda_\alpha} e^{-\lambda_\alpha} \sum_{n=1}^{\infty} \frac{n^2 \, I_n(\lambda_\alpha)}{\omega^2 - n^2 \, \Omega^2_\mathrm{c,\alpha}} = 0 \quad .
	\label{eqn:disp_rel_full_bernstein}
\end{equation}

This expression uses modified Bessel functions $I_n$ of the first kind. For
frequencies on the order of the electron gyro frequency and higher, the ions can
be considered stationary and all terms connected to them can be
dropped\footnote{A more careful treatment finds that each mode resulting from
the dispersion relation Eq.~\eqref{eqn:disp_rel_electron_bernstein} actually
consists of a large number of modes separated by multiples of the ion gyro
frequency. Resolving this stack of modes is however complicated both in
experiments and simulations.}. The dispersion relation then simplifies and
reads:

\begin{equation}
	1 - \frac{2 \, \Omega_\mathrm{c,e}^2}{k^2 \, \lambda_\mathrm{D}^2} \, e^{-\lambda_\mathrm{e}} \sum_{n=1}^{\infty} \frac{n^2 \, I_n(\lambda_\mathrm{e})}{\omega^2 - n^2 \, \Omega_\mathrm{c,e}^2} = 0\quad.
	\label{eqn:disp_rel_electron_bernstein}
\end{equation}

It is also possible to consider Bernstein waves at lower frequencies. At ion
length scales of $\lambda_\mathrm{i} \approx 1$ we find that $\lambda_e \approx
m_\mathrm{e}/m_\mathrm{i} \lambda_\mathrm{i} \ll 1$. Given that the Bessel
function of order $n > 1$ vanish for small arguments, this implies that we can
drop the electron terms when considering ion scales. In this limit
Eq.~\eqref{eqn:disp_rel_full_bernstein} can be approximated by:

\begin{equation}
	1 - \frac{\omega_\mathrm{p,e}^2}{\Omega_\mathrm{c,e}^2} - \frac{2 \, \Omega_\mathrm{c,i}^2}{k^2 \, \lambda_\mathrm{D}^2} \, e^{-\lambda_\mathrm{i}} \sum_{n=1}^{\infty} \frac{n^2 \, I_n(\lambda_\mathrm{i})}{\omega^2 - n^2 \, \Omega_\mathrm{c,i}^2} = 0\quad.
	\label{eqn:disp_rel_ion_bernstein}
\end{equation}

The term containing electron quantities is not present in the same way for
electron Bernstein waves and represents a shielding effect from the electrons.

For high orders of $n$ the simplification of representing the electrons by a
constant term gets increasingly wrong, especially if the mass ratio
$m_\mathrm{i} / m_\mathrm{e}$ is not sufficiently large. However, as can be
seen from the parameters in Tab.~\ref{tab:ion_bernstein}, this is not a problem
for the test case presented here.

\section{Numerical Implementation}
\label{sec:implementation}

Most of the simulations that were performed to produce the illustrations in
this paper used the PiC Code ACRONYM \cite{Kilian_2012} or extensions thereof.
This simulation code is quite flexible and implements a wide range of plasma
models. The simulation domain can be simulated with one, two or three
spatially resolved dimensions (fields and velocities are always represented by
three component vectors) and their boundaries can be periodic, reflecting or
absorbing.  For the test case, only one spatially resolved dimension and
periodic boundary conditions are used to reduce numerical effort and
implementation requirements.

Kinetic species are represented by macro particles. Their charge and current
density is deposited onto the grid using second-order interpolation with the
triangular-shaped-cloud (TSC) shape function. The current is deposited based on
the same shape function using $q\,\vec{v}$ for one-dimensional simulations or
with the charge-conserving method of \cite{Esirkepov_2001} for two- and
three-dimensional simulations.

Electrons in the plasma can be represented either in this way for fully kinetic
simulations, or implicitly as a fluid with or without electron inertia, following the
Electron Magneto-Hydro-Dynamic (EMHD) solver of \cite{Jain_2003}.

\subsection{Electromagnetic}

In electromagnetic simulations, the electric and magnetic field is evolved from
the homogeneous initial state using the Finite-Difference-Time-Domain (FDTD) method:

\begin{equation}
	\begin{array}{lclcl}
		\vec{B}^\mathrm{\;t+1/2} &=& \vec{B}^\mathrm{\;t-1/2} &-& \mathrm{c} \cdot \Delta t \cdot \nabla \times \vec{E}^\mathrm{\;t}, \\
		\vec{E}^\mathrm{\;t+1}   &=& \vec{E}^\mathrm{\;t}     &+& \mathrm{c} \cdot \Delta t \cdot \nabla \times \vec{B}^\mathrm{\;t+1/2} - 4 \pi \Delta t \cdot \vec{\jmath}^\mathrm{\;t+1/2}.
	\end{array}
	\label{eqn:maxwell_disc}
\end{equation}

The field quantities are stored in a staggered grid following the idea by
\cite{Yee_1966}, which allows for a very straight-forward calculation of the
curl that is accurate to second order. This field solver leads to a
modification of the dispersion relation at large $\omega$ and $k$. In the case
of the electromagnetic mode propagating along one axis of the spatial grid, the
resulting dispersion relation is
\begin{equation}
	\left(\frac{2}{\Delta t}\right)^2 \sin^2\left(\frac{\omega \; \Delta t}{2}\right) = \omega_\mathrm{p}^2 + \mathrm{c}^2 \; \left(\frac{2}{\Delta x}\right)^2 \sin^2\left(\frac{k \; \Delta x}{2}\right)\quad.
	\label{eqn:numerical_disp_em_em}
\end{equation}
For frequencies that are not close to the respective Nyquist limit, the
modification is negligible.

\subsection{Radiation-free}

The numerical implementation of the radiation-free model is harder than
Eqs.~\eqref{eqn:maxwell_darwin1}-\eqref{eqn:maxwell_darwin3} make believe at first.
The reason is the fact that $\vec{E}_\mathrm{T}$ depends on the time derivative
of the transverse component $\vec{\jmath}_\mathrm{T}$ of the current. The
change in current, however, is of course connected to the acceleration of the
particles which in turn depends on the electric field. An overly naive time
discretization is therefore violently unstable. Our code follows the method of
\cite{Decyk_2011}, which solves the problem in the following way.

In the first part of a time step, the charge density $\rho$ is deposited onto the
grid and Eq.~\eqref{eqn:maxwell_darwin3} is solved with a Fourier based solver to
get $\vec{E}_\mathrm{L}$.

The new values of $\vec{E}_\mathrm{T}$ and $\vec{B}$ are calculated using the
following iterative scheme: First $\vec{\jmath}$ and
$\partial\vec{\jmath}/\partial t$ are deposited onto the grid using the
assumption that the particle velocities remain unchanged.

Once the current contributions are known, it is possible to solve for the field
components.  And as soon as all fields are known, a better prediction of the
particle velocities can be made. Using the better estimate for the particle
velocities, a better estimate of the current can be deposited onto the
grid. This, in turn, allows for a refinement of the field components.

The iteration scheme converges quickly, after two or three iterations. At this
point, the particle velocity can be updated based on the best current prediction
for the new velocity and the code can proceed to the next time step.

\subsection{Electrostatic}

The electrostatic model uses a spectral solver to calculate the longitudinal
electric field. This solver makes use of the fact that the charge density can
be replaced by its Fourier series
\begin{equation}
    \rho\left(\vec{r}\right)= \sum_{kx,ky,kz} \tilde\rho(\vec{k}) \exp\left(\imath k_x x + \imath k_y x + \imath k_z z\right)
\end{equation}
in Eq.~\eqref{eqn:maxwell_darwin3}. The components of the electric field can be
rewritten in the same way and the derivative can act directly on the
exponential functions. As the different Fourier modes are orthogonal, the
resulting equation has to be satisfied for each mode separately, which leads to
the following relation between the charge density and the electric field in the
spectral domain:

\begin{equation}
	\tilde{\vec{E}}(\vec{k}) = - 4 \pi \, \imath \, \frac{\tilde\rho(\vec{k}) \, \vec{k}}{|\vec{k}|^2} \quad .
\end{equation}

This way, the electric field can be calculated by performing a Fourier
transform on $\rho$, multiplying every component in $k$-space with
$ - 4 \pi \, \imath \, \vec{k} |\vec{k}|^{-2}$, which can be considered a
convolution with the Green function of free space and transforming the
resulting field back to real space.

Both the radiation-free model and the electrostatic model have a gap at $k = 0$
in the spectrum of the electric field. This is an artifact of solving Poisson's
equation in the Fourier domain. The only possible source term that would
produce $\tilde{E}(k =  0)$ is $\tilde{\rho}(k = 0)$. This quantity, however, is
the net charge density which vanishes when averaging over the entire simulation
domain that contains an equal number of positive and negative charged particles.

\subsection{Vlasov-Hybrid Simulation}
\label{pp:vhs_description}

We also used a second independent implementation of the electrostatic plasma
models, which is not based on the PiC method, but instead follows the
Vlasov-Hybrid-Simulation (VHS) method by \cite{Nunn_1993}. This method is not a
hybrid between a kinetic and a fluid part, but a hybrid between an Eulerian
description of phase space density and a Lagrangian description using macro
particles. Where a PiC code represents chunks of phase space density as macro
particles of constant weight and deposits their charge or current onto a grid,
a VHS code reconstructs the phase space density on a grid. It then integrates
out the velocity direction(s) to obtain moments of the distribution function
such as charge density. The reconstruction step requires extra effort but
allows for the use of macro particles with significantly different weights
without losing the effect of markers that represent low phase space density.
This makes VHS a technique with a very low level of numerical noise. An open
source implementation and a description of the code have been submitted for
publication elsewhere.

\subsection{Implicit Electron Fluid}

Details of the hybrid code is decribed in \cite{Munoz_2016}. On a high level it
computes the time evolution of the ions, just as a PiC code would and
determines the ions charge density $n_\mathrm{i}$ and current density
$\vec{\jmath}_\mathrm{i}$. Using those the generalized vorticity can be updated
as described in Eq.~\eqref{eqn:vorticityupdate}. After that, the magnetic field
$\vec{B}$ can be computed from a version of Eq.~\eqref{eqn:vorticitydef}, where
the electron flow speed has been eliminated using
Eq.~\eqref{eqn:currentbalance}. Then the electric field $\vec{E}$ can be
determined from the generalized Ohm's law given in
Eq.~\eqref{eqn:emhd_electric_field}.

\subsection{Linearized Dielectric Tensor}

To verify the analytically known plasma modes and to check for thermal effects
that are neglected in their cold plasma description, we also used the WHAMP code
(see \cite{Roennmark_1982}). This code does not evolve the full plasma model but
starts with the linearization of the time-independent equations. Assuming a
parametrized velocity distribution function it uses analytic expressions to
approximate the dielectric tensor of a warm multi-component plasma. The plasma
dispersion function that is needed in that description is numerically
approximated by a Pad\'e approximation and the resulting tensor is solved
numerically using Newton iteration.

\section{Test Problems}
\label{sec:test}

It is not easy to choose plasma parameters that are accessible to a large range
of simulation codes and that produce results that can be compared in a
meaningful manner. This can be seen, for example, in the case of the thermal
speed of electrons. Very small values -- and consequently low temperatures --
make the comparison with predictions for cool or cold plasmas easier. Explicit
PiC codes, on the other hand, have small time steps that are set by the speed
of light. If thermal particles move at a tiny fraction of the speed of light,
they only move a tiny distance per time step and the code has to compute many
time steps. As a compromise and to avoid relativistic effects, an electron
thermal speed of five percent of the speed of light was chosen. All three
initial velocity components of the electrons are drawn independently from a
normal distribution with zero mean and standard deviation $v_\mathrm{th,e}$.
Using the relation $m_\mathrm{e} v_\mathrm{th,e}^2 = k_\mathrm{B}
T_\mathrm{e}$, we determine the equivalent temperature of 14.79 MK.

The plasma contains protons (single positive charge, natural mass ratio unless
specifically noted otherwise) as neutralizing (ion) species.  The ion
temperature $T_\mathrm{i}$ is set equal to the electron temperature, therefore,
their thermal speed is lower by a factor of $\sqrt{m_\mathrm{i} /
m_\mathrm{e}}$.

The absolute value of the plasma frequency has no such direct physical
implications. However, it sets the Debye length and thereby the size of the
grid cells. A value of $10^9\, \mathrm{rad}/\mathrm{s}$ was chosen for the
electrons, which corresponds to a density of $3.14 \cdot 10^8$ particles per
cubic centimeter. The protons have the same density to fulfill charge
neutrality, which translates to a proton plasma frequency of $2.33 \cdot 10^7$
rad/s. The contribution of the protons to the total plasma frequency is
negligible.

Using this temperature and frequency scale, the Debye length turns out to be
1.497~cm. To avoid grid heating and other numerical effects, the cell size of
each grid cell should be slightly smaller (see e.g. \cite{Birdsall_2005}),
which suggests the round value of 1~cm.

The only physical parameter that still needs to be specified is the magnetic
field strength. Here we select 2.843~mT which corresponds to a high density
plasma with $2 \Omega_\mathrm{c,e} = \omega_\mathrm{p,e}$. Tab.~\ref{tab:baseline_def} lists
all the defining parameters in compact form and Tab.~\ref{tab:baseline_res}
has some other derived plasma parameters.

\begin{table}[hbpt]
\begin{center}
\begin{tabular}{l l r}
electron plasma frequency & $\omega_\mathrm{p,e}$ & $1.000 \cdot 10^9\, \mathrm{rad} / \mathrm{s}$ \\
electron gyro frequency & $\Omega_\mathrm{c,e}$ & $5.000 \cdot 10^8\, \mathrm{rad} / \mathrm{s}$ \\
electron thermal speed & $v_\mathrm{th,e}$ & 0.050 c \\
mass ratio & $m_\mathrm{i} / m_\mathrm{e}$ & 1836 \\
\end{tabular}
\end{center}
\caption{Common choice of simulation parameters to be used in all simulations unless noted otherwise.}
\label{tab:baseline_def}
\end{table}

\begin{table}[hbpt]
\begin{center}
\begin{tabular}{l l r}
ion thermal speed & $v_\mathrm{th,i}$ & $\approx$ 0.001 c \\
temperature & $T$ & 14.79 MK\\
            &     & 1275 eV\\
Debye length & $\lambda_\mathrm{D}$ & 1.497 cm \\
grid size & $\Delta x$ & 1.000 cm\\
electron gyro radius & $r_\mathrm{e}$ & 2.998 cm \\
magnetic field & $B_0$ & 2.843 mT\\
               &       & 28.43 G\\
Alfv\'en speed & $v_\mathrm{A}$ & 0.012 c
\end{tabular}
\end{center}
\caption{Resulting plasma parameters based on the choice made in Tab.~\ref{tab:baseline_def}.}
\label{tab:baseline_res}
\end{table}

Some of the simulations that depend on one spatial dimension were actually
run with a 3d PiC code with 4 cells width and periodic boundary conditions in
the negligible directions. Each cell contains eight computational macro
particles per species.

\begin{figure}[htbp!]
\begin{center}
	\includegraphics[width=0.8\columnwidth]{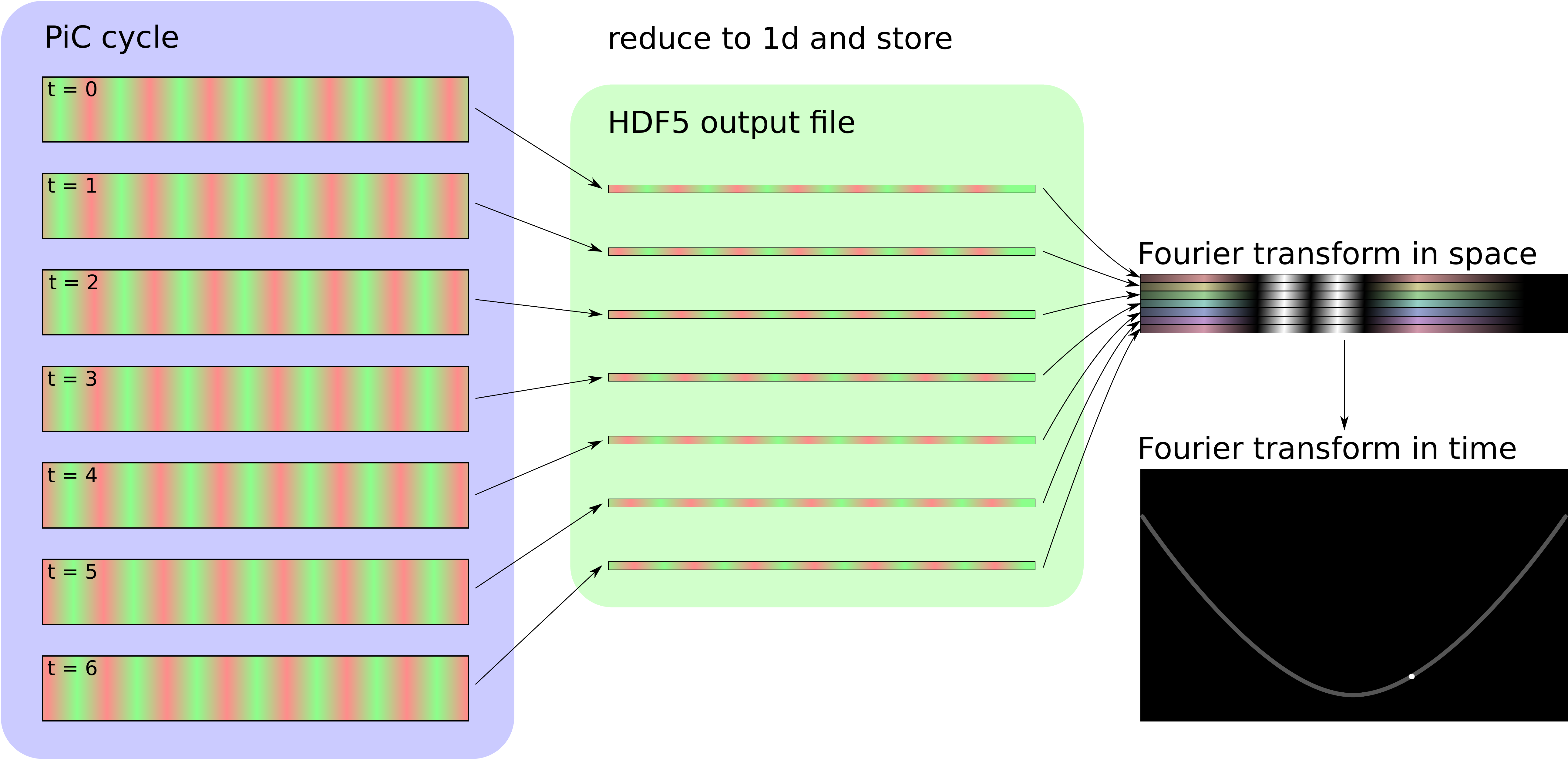}
	\caption{Sketch of the analysis pipeline.}
	\label{fig:spectral_analysis}
\end{center}
\end{figure}

After the numerical simulation has been performed, it is necessary to extract
properties of the wave modes and compare with the expected behavior. To this
end it is very useful to plot the energy density of a field component as a
function of $k$ and $\omega$. The energy density is sharply localized along
linear wave modes and characteristic frequencies (e.g. the low frequency cutoff
of the electromagnetic mode) can easily and reliably be extracted.

Fig.~\ref{fig:spectral_analysis} shows a sketch of the analysis that is
performed. For the following explanation, we assume that we are interested in
one component of the transverse electric field, other field components work
analogously. Depending on the simulation code, $E_\mathrm{x}(z,t)$ or possibly
$E_\mathrm{x}(x,y,z,t)$ is computed in every time step. This quantity is stored
along with the necessary meta-data in a HDF5 file for later analysis.

As I/O can be a bottleneck for the simulation code, it is desirable to reduce
overall output requirements. For the study of low frequency wave modes, it is
usually sufficient to perform output every $N_\mathrm{io}$ time steps as long
as
\begin{equation}
	\frac{\pi}{N_\mathrm{io}\,\Delta t} = \frac{\pi}{\Delta t_\mathrm{io}} < \omega
\end{equation}
holds for the highest frequency $\omega$ that is of interest. Similarly, it
should be considered whether the code can average over the negligible directions
before performing output.

In post-processing, $E_\mathrm{x}(z,t)$ is Fourier transformed in space and
time to yield $\tilde{E}_\mathrm{x}(k_z,\omega)$. To scale the axis correctly,
it is useful to know that the frequency range one gets out of snapshots that
are separated by $\Delta t_\mathrm{io}$ is $0 \dots \pi / t_\mathrm{io}$ in
steps of $\pi / T_\mathrm{sim}$. The range in $k$ is $- \pi / \Delta x \dots
\pi \Delta x$. In most cases it is advisable to rescale from the units used in
the code -- CGS with values in in 1/s and 1/cm for the codes used here -- to
plasma scales such as $\omega_\mathrm{p}$, $\Omega_\mathrm{c,e}$ before
plotting.

\FloatBarrier
\subsection{Test 1: Electromagnetic Mode}
\label{subsec:test:electromagnetic}

The cheapest test is designed to capture the electromagnetic mode. It does not
use any background magnetic field and the size is given in Tab.~\ref{tab:em_par}.

\begin{table}[hbpt]
\begin{center}
\begin{tabular}{l l r}
simulation domain & $L_z$ & $2048 \, \Delta x$ \\
simulation duration & $T_\mathrm{sim}$ & $200 \, \omega_{p,e}^{-1}$ \\
& & $10400 \, \Delta t$
\end{tabular}
\end{center}
\caption{Simulation size to study the electromagnetic mode.}
\label{tab:em_par}
\end{table}

\begin{figure}[htbp!]
\begin{center}
	\includegraphics[width=\columnwidth]{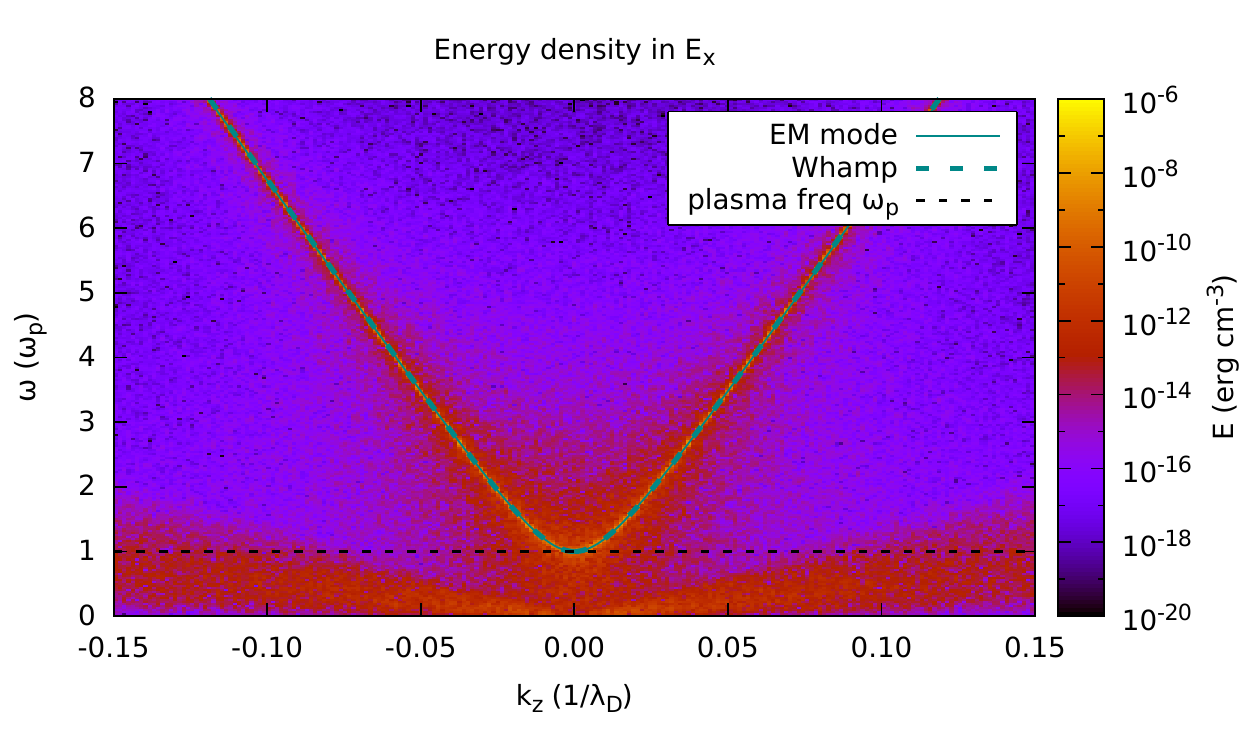}
\end{center}
\caption{Energy density in one transverse component of the electric field as
simulated by the electromagnetic plasma model. The electromagnetic mode is
clearly visible, including the cutoff at the plasma frequency
$\omega_\mathrm{p}$. The simulation parameters can be found in
Tabs.~\ref{tab:baseline_def} and \ref{tab:em_par}, but the simulation is
performed without a magnetic field.}
\label{fig:disp_em_em}
\end{figure}

Fig.~\ref{fig:disp_em_em} shows that the energy is mostly concentrated along
the dispersion relation predicted by cold plasma theory (given by
Eq.~\eqref{eqn:disp_em_em}, shown as a dashed line) and has a cutoff at the
plasma frequency $\omega_\mathrm{p}$, which is indicated by the horizontal
dashed line. Predictions for the absolute magnitude and the low frequency noise
are given in \cite{Sitenko_1967}, but are not discussed here.

This mode is the cheapest way to determine the plasma frequency from the
simulated data and compare it to the desired value from the simulation input.
Many numerical problems (wrong normalization, errors in the charge or current
deposition, problems in the particle pusher) can alter this mode, so it is
ideally suited as a quick regression check after code modifications.

\begin{figure}[htbp!]
\begin{center}
	\includegraphics[width=\columnwidth]{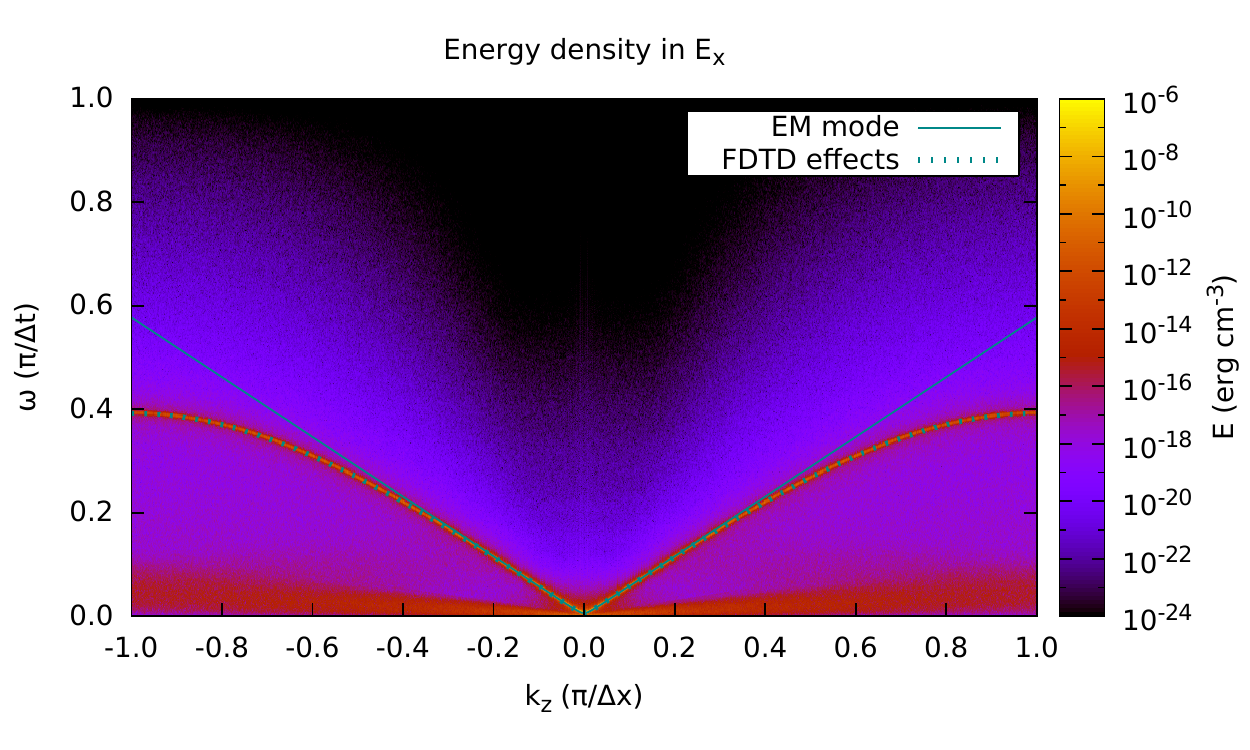}
\end{center}
\caption{Spectral energy density up to the resolution limits permitted by the
finite cell size and time step length. This plot uses data from the same
simulation as Fig.~\ref{fig:disp_em_em}. At high frequencies and large $k$ close
to $\pm \pi / \Delta x$, significant numerical dispersion is visible.}
\label{fig:disp_em_em_lz}
\end{figure}

At large $k$ -- closer to the Nyquist limit imposed by the grid size $\Delta x$
-- numerical dispersion effects occur that result from the discrete Maxwell solver in
the simulation code. Fig.~\ref{fig:disp_em_em_lz} shows this effect of the FDTD
algorithm that is used to solve Maxwell's equations in time. The modified
dispersion relation is given by Eq.~\eqref{eqn:numerical_disp_em_em}, in good
agreement with our results.

\begin{figure}[htbp]
\begin{center}
	\includegraphics[width=\columnwidth]{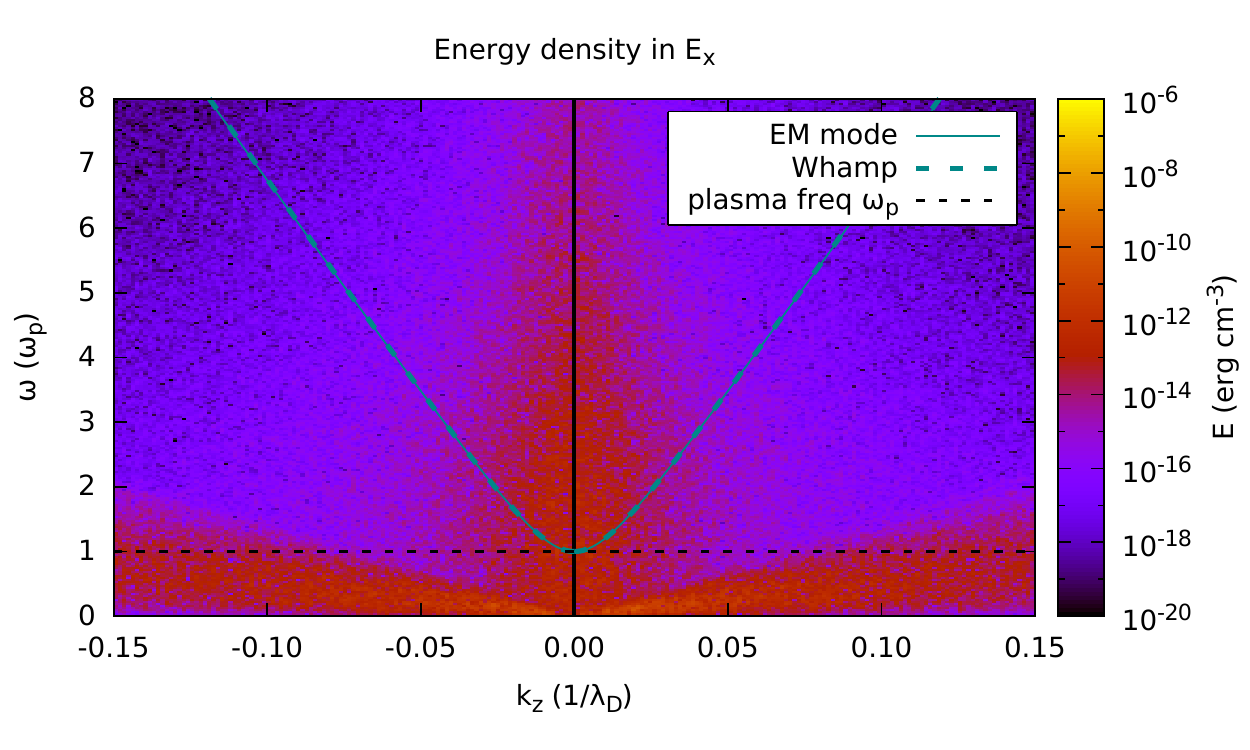}
\end{center}
\caption{Energy density in one component of the radiation-free plasma model.
For this plot the same parameters listed in Tabs.~\ref{tab:baseline_def} and
\ref{tab:em_par} were used, with the exception of the background magnetic
field. Compared to Fig.~\ref{fig:disp_em_em}, the electromagnetic mode is
missing.}
\label{fig:darwin_em}
\end{figure}

In the radiation free plasma model, the electromagnetic mode is explicitly
removed. The remaining part of the transverse spectrum is shown in
Fig.~\ref{fig:darwin_em}. No well defined mode is left in the unmagnetized
case. At small $\omega$ and low phase velocities, a diffuse mode is visible
that can be identified as ion acoustic waves. These, however, have a broadband
spectrum without sharp features that would provide a good test problem.

The increase in noise at low $k$ can be attributed to a well known numerical
instability at scales larger than the electron skin depth in the radiation-free
plasma model. This instability can be removed through the introduction of a
shift constant in the equation for the transverse electric field, but some
noise remains (see \cite{Decyk_2011} for details).

\FloatBarrier
\subsection{Test 2: High frequency L and R Modes}
\label{subsec:test:hf_lr}

The addition of a background magnetic field (of 2.843 mT in this test problem)
along the $z$ direction splits the electromagnetic mode into two modes with
circular polarization.

To study polarization properties of the waves, one switches from a standard
Cartesian basis (with transverse components $\hat{x}, \hat{y}$) to a circular
basis. In plasma physics phase convention, the new basis vectors are given by

\begin{align}
	\hat{l} &= \frac{1}{\sqrt{2}} \left(\hat{x} - \imath\, \hat{y}\right) \quad ,\\
	\hat{r} &= \frac{1}{\sqrt{2}} \left(\hat{x} + \imath\, \hat{y}\right) \quad .
\end{align}

Instead of performing the analysis that is sketched in
Fig.~\ref{fig:spectral_analysis} on a field component in the Cartesian basis
(e.g.  $E_\mathrm{x}(z,t)$), it is possible to combine $E_\mathrm{x}(z,t)$
and $E_\mathrm{y}(z,t)$ in the following way:
\begin{equation}
	E_\mathrm{l,r} = \frac{1}{\sqrt{2}} \left(E_\mathrm{x}(z,t) \mp \imath \, E_\mathrm{y}(z,t)\right) \quad .
	\label{eqn:def_circ_electric}
\end{equation}

\begin{figure}[htbp!]
\begin{center}
	\includegraphics[width=\columnwidth]{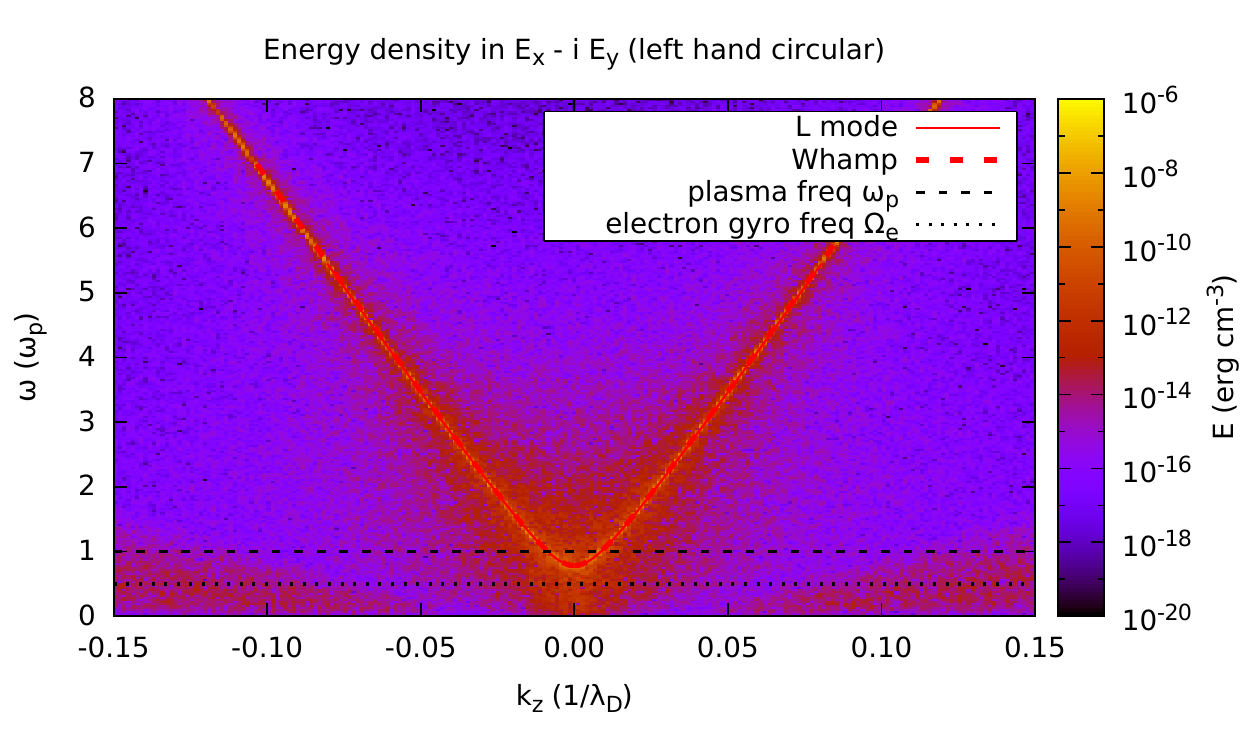}
\end{center}
\caption{Energy density in the left handed circularly polarized component of
the electric field produced by the electromagnetic plasma model. This plot is
based on parameters given in Tabs.~\ref{tab:baseline_def} and \ref{tab:em_par}.
The spectral energy density is concentrated along the high frequency branch of
the L mode, propagating along the background magnetic field.}
\label{fig:disp_em_hf_l}
\end{figure}

\begin{figure}[htbp!]
\begin{center}
	\includegraphics[width=\columnwidth]{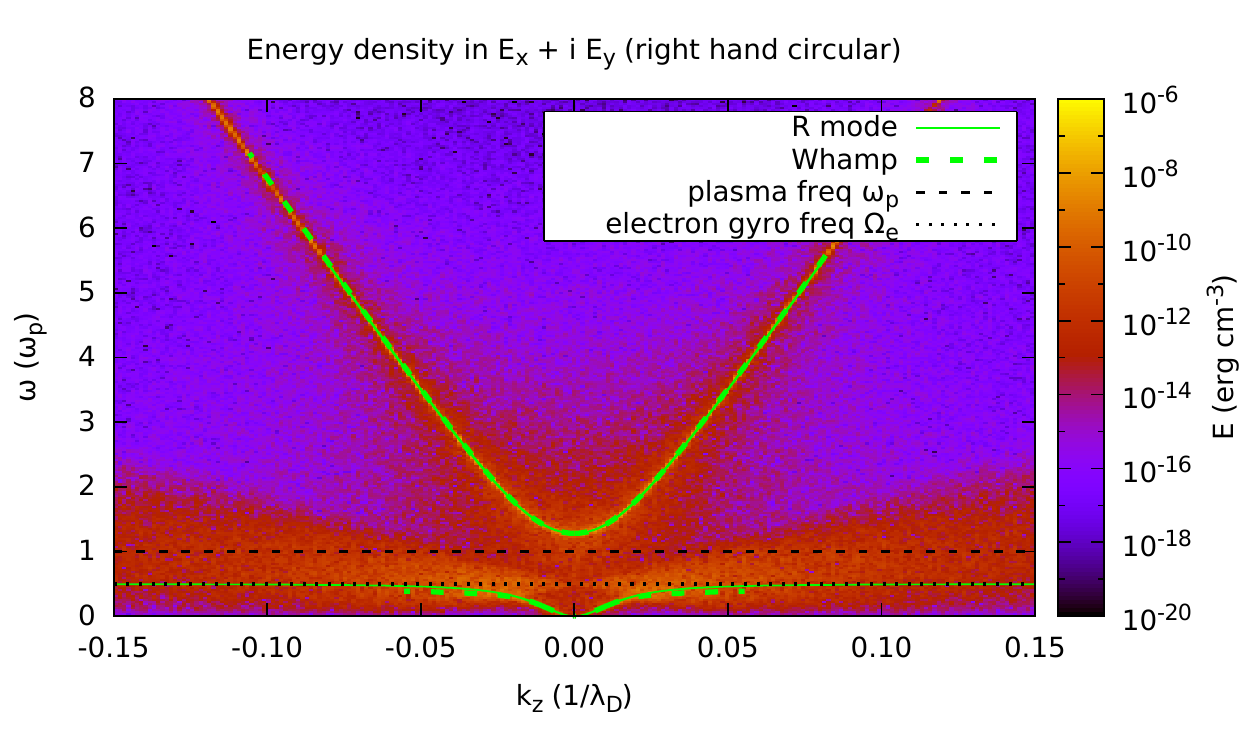}
\end{center}
\caption{Energy density in the right handed circularly polarized component of
the electric field produced by the electromagnetic plasma model. This plot uses
the same parameters as Fig.~\ref{fig:disp_em_hf_l}, but the simulation results
are combined differently to display the other circular polarization basis. Both
high and low frequency branches of the R mode are visible.}
\label{fig:disp_em_hf_r}
\end{figure}

As usual, a Fourier transform\footnote{This requires a complex-to-complex
transform. However, in our analysis pipeline we use such a transformation, at
slightly higher computational cost, even for purely real input data to simplify
memory management. Thus the analysis in circular basis introduces no significant
extra complication.} in space and time is performed to yield
$\tilde{E}_\mathrm{l,r}(k_\mathrm{z},\omega)$. Figs.~\ref{fig:disp_em_hf_l} and
\ref{fig:disp_em_hf_r} how the spectral energy density
$|\tilde{E}_\mathrm{l,r}|^2$ respectively.

In the circular basis only the wave modes with the matching circular
polarization appear, thus confirming that the numerical implementation
reproduces the expected polarization properties. Both modes follow the
analytically predicted dispersion relation given in Eq.~\eqref{eqn:disp_em_lr}.
The cutoff is shifted (away from the cutoff at $\omega_\mathrm{p}$ in the
unmagnetized case) by about $\pm 1/2 \, \Omega_\mathrm{c,e}$ as expected.

In the right hand polarization, shown in Fig.~\ref{fig:disp_em_hf_r}, two
additional features are visible. One is a low frequency component with a
resonance at $\Omega_\mathrm{c,e}$, which will be studied in more detail in a
following test. The other feature is a triangular region of fluctuations that is
caused by gyrating electrons. Within that region that is centered on
$\Omega_\mathrm{c,e}$ and bounded by approximately $\pm 3 v_\mathrm{th,e} \, k$,
the dispersion relation of the R mode is modified. At larger $k$ the wave is
absorbed. Both effects are captured by WHAMP and studied in more detail in
Test~6.

\FloatBarrier
\subsection{Test 3: Extraordinary Mode}
\label{subsec:test:extraordinary}

Keeping the background magnetic field along $z$ and rotating the simulation box
to point along $x$, allows to study waves that propagate across the magnetic
field. (Alternatively, one can change the direction of the magnetic field, in
which case field components switch behavior.)

\begin{figure}[htbp!]
\begin{center}
	\includegraphics[width=\columnwidth]{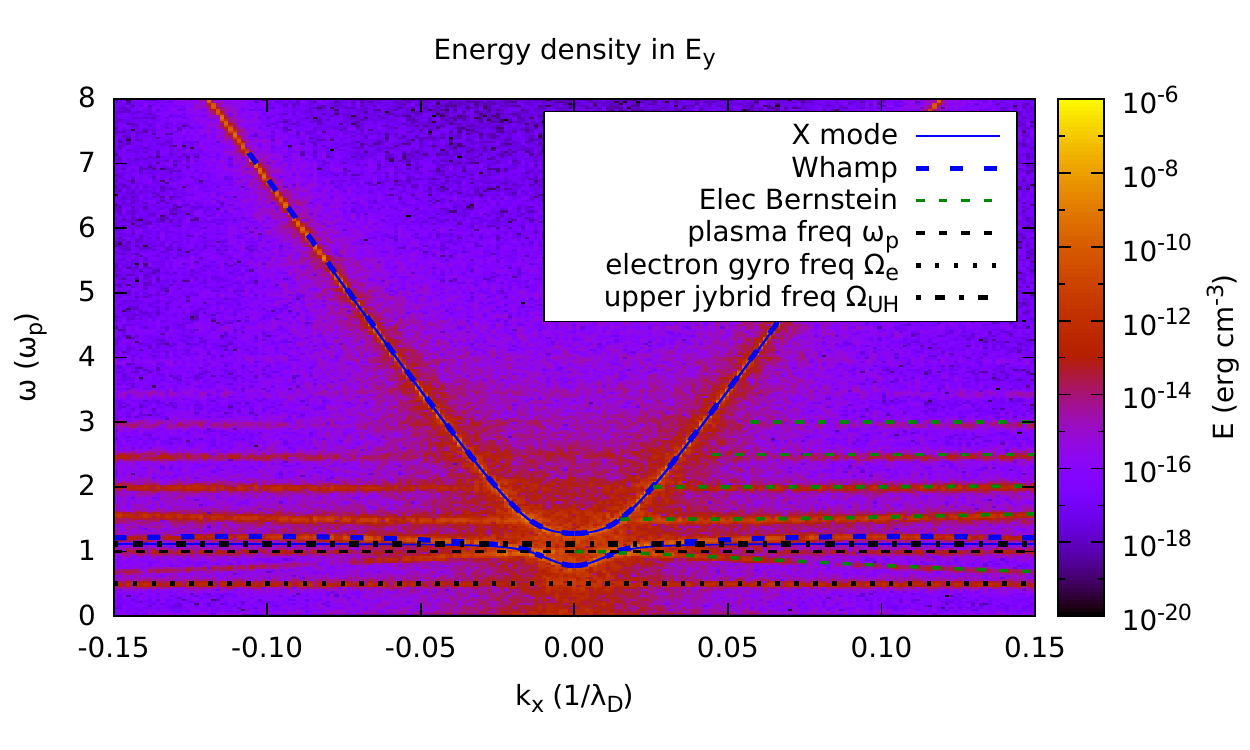}
\end{center}
\caption{Energy density in the electric field component that is perpendicular
to both the background magnetic field and the propagation direction. The plot
shows simulation results from the electromagnetic plasma model with parameters
given in Tabs.~\ref{tab:baseline_def} and \ref{tab:em_par}. Both branches of the
X mode are clearly visible. Due to the finite temperature, harmonics of the
electron gyro frequency and the first few electron Bernstein modes are also
visible.}
\label{fig:disp_em_x}
\end{figure}

As expected, the field component perpendicular to $k$ and $B_0$ shows the
extraordinary mode. Both the high frequency branch above the upper hybrid
frequency and the branch close to the plasma frequency show the expected
dispersion relation given in Eq.~\ref{eqn:disp_x_mode}. Thermal effects are not
important for this mode, as can be seen by the excellent agreement between the
predictions of cold plasma theory and the numerical solution by WHAMP that
includes finite temperature effects\footnote{In some sense this method is the
opposite of the quiet start method described in \cite{Birdsall_2005} that aims
to initialize a thermal distribution with as little fluctuations as possible}.

Fig.~\ref{fig:disp_em_x} also shows approximately horizontal features at
harmonics of the electron gyro frequency. These are due to electron Bernstein
waves which  can only be explained by the thermal effects and are studied in
more detail in Test~7.

\FloatBarrier
\subsection{Test 4: Langmuir Mode}
\label{subsec:test:langmuir}

This test problem focuses on longitudinal waves which, in the unmagnetized
case, are represented by the Langmuir mode. As mentioned previously, this mode
is a result of plasma oscillations in a plasma of finite temperature.
Consequently, it can be used to determine the thermal speed of the electrons
$v_\mathrm{th,e}$ and thereby the temperature $T_\mathrm{e}$. This is of
interest in codes that start with all particles at rest and rely on the initial
fluctuations in the charge density to generate a thermal velocity distribution.

\begin{figure}[htbp]
\begin{center}
	\includegraphics[width=\columnwidth]{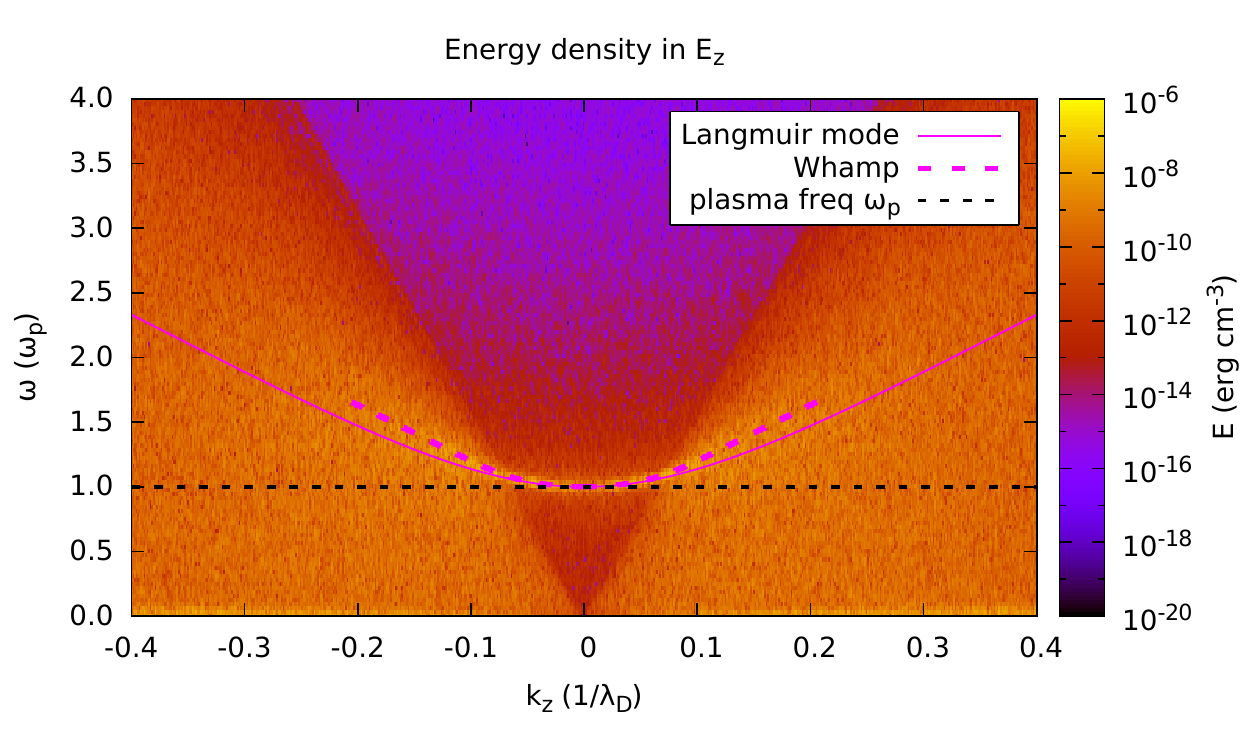}
\end{center}
\caption{Energy density in the longitudinal component of the electric field.
The simulation was performed with using the electromagnetic plasma model and
the parameters given in Tabs.~\ref{tab:baseline_def} and \ref{tab:em_par}, but
without background magnetic field. At not too large $k$, the Langmuir mode is
clearly visible.}
\label{fig:em_es}
\end{figure}

Fig.~\ref{fig:em_es} compares the energy distribution in the longitudinal
electric field with the expected dispersion relation of a Langmuir mode in a
plasma of the same temperature that was used to generate the initial velocity
distribution. For small $k$, the agreement with the analytic prediction is very
good. At intermediate $k$, there are deviations from the analytic prediction
due to the rather large electron temperature. WHAMP, however, is able to
accurately predict the behavior of the plasma. At large $k$, the wave is
strongly damped and not visible in the spectral energy distribution. This
effect is also predicted by WHAMP, but missing from the analytical dispersion
relation given in Eq.~\eqref{eqn:disp_langmuir}.

\begin{figure}[htbp]
\begin{center}
	\includegraphics[width=\columnwidth]{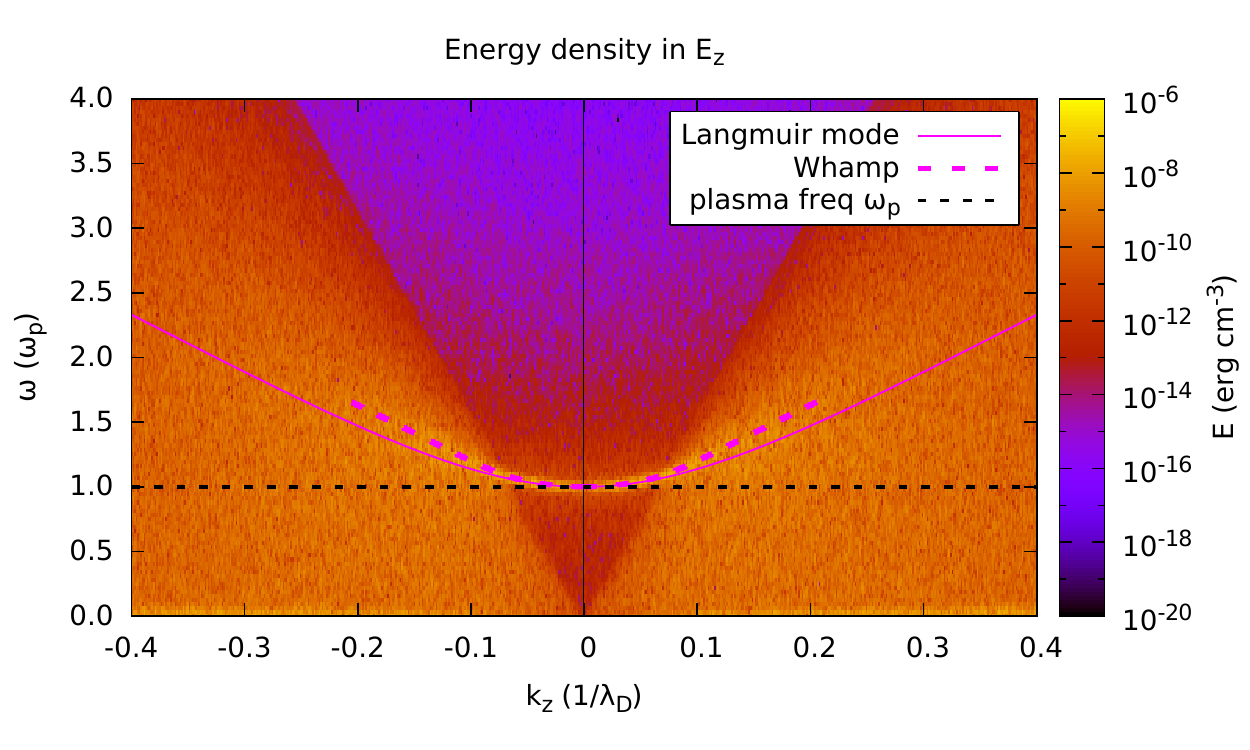}
\end{center}
\caption{Energy density in the longitudinal component of the electric field.
The setup and parameters are identical to Fig.~\ref{fig:em_es} but the electric
field is computed by the spectral solver that is used in both the
radiation-free and the electrostatic plasma model.}
\label{fig:es_es}
\end{figure}

Fig.~\ref{fig:es_es} shows the longitudinal electric field, as determined by the
spectral solver that is used in the radiation-free plasma model as well as the
electrostatic plasma model in ACRONYM. The gap at $k = 0$ is an artifact of the
spectral solver that was mentioned before. At all other $k$ the match to the
electromagnetic plasma model and the prediction from theory is very good.

\begin{figure}[htbp]
\begin{center}
	\includegraphics[width=\columnwidth]{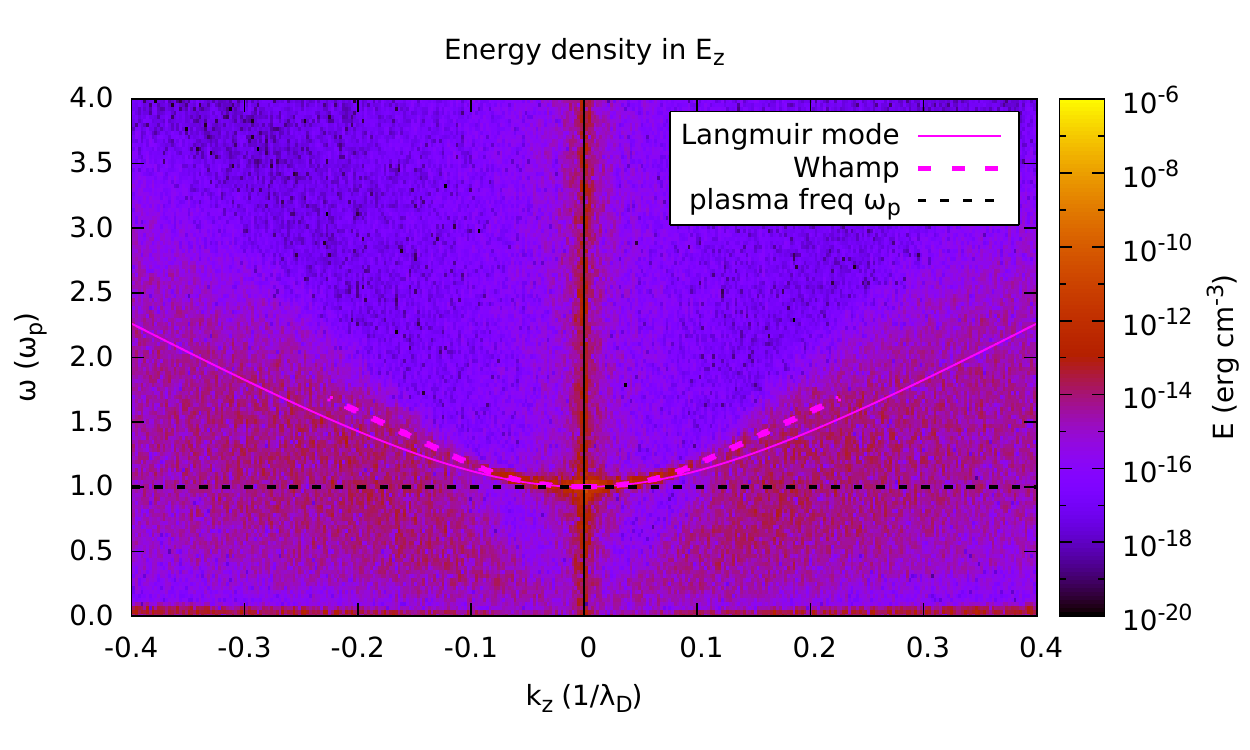}
\end{center}
\caption{Plot of longitudinal perturbations in the electric field for the
alternative implementation of the electrostatic plasma model using the VHS
technique. See page~\pageref{pp:vhs_description} for a description and
Tabs.~\ref{tab:baseline_def} and \ref{tab:em_par} for parameters.}
\label{fig:vhs_es}
\end{figure}

Fig.~\ref{fig:vhs_es} shows the longitudinal electric field for the alternative
implementation of the electrostatic plasma model using the VHS technique. This
method has a very low level of intrinsic noise. To make the Langmuir mode
visible, the initial density of each species was randomly perturbed on every
point of the phase space grid by plus or minus five percent. This reproduces
the Langmuir mode at about the same strength as it appears in the PiC
simulations.

Very visible at least in Fig.~\ref{fig:em_es} and \ref{fig:es_es} and still
recognizable in Fig.~\ref{fig:vhs_es} is the change in the broadband noise
floor that is caused by thermal electrons and reaches up to
\begin{equation}
	\omega = 3 \sqrt{2}\, v_\mathrm{th,e}\, k \quad .
\end{equation}
The reduction of this noise floor by at least four orders of magnitude is the
main reason to consider implementing the electrostatic plasma model using the
VHS method.

The hybrid plasma models do not include kinetic effects of the electrons and
assume instantaneous neutrality which, therefore, removes the Langmuir mode as
well as the thermal broadband noise.

\FloatBarrier
\subsection{Test 5: Bernstein Modes}
\label{subsec:test:electron_bernstein}

\begin{figure}[htbp]
\begin{center}
	\includegraphics[width=\columnwidth]{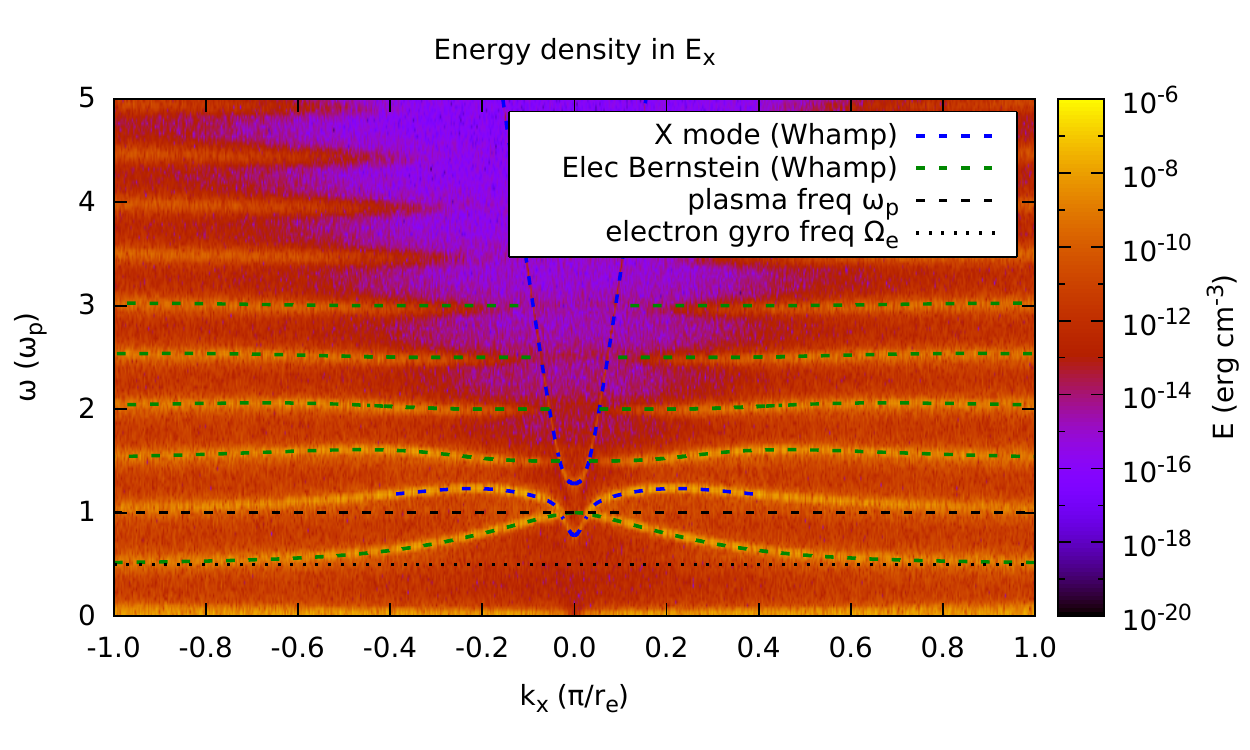}
\end{center}
\caption{Energy density in the longitudinal component of the electric field.
The simulation was performed using the electromagnetic plasma model, using
parameters from Tabs.~\ref{tab:baseline_def} and \ref{tab:em_par}. The first few
electron Bernstein modes are clearly visible.}
\label{fig:em_bernstein}
\end{figure}

Fig.~\ref{fig:em_bernstein} shows the electron Bernstein modes described in
Sec.~\ref{subsec:wave:electronbernstein}. A comparison with theoretical
predictions is hampered by the complicated dispersion relations of these modes.
Using the leading terms of the infinite sums, it is possible to plot the first
few modes. The figure also contains some influence from the X mode that has a
longitudinal components at low $k$.

\begin{figure}[htbp]
\begin{center}
	\includegraphics[width=\columnwidth]{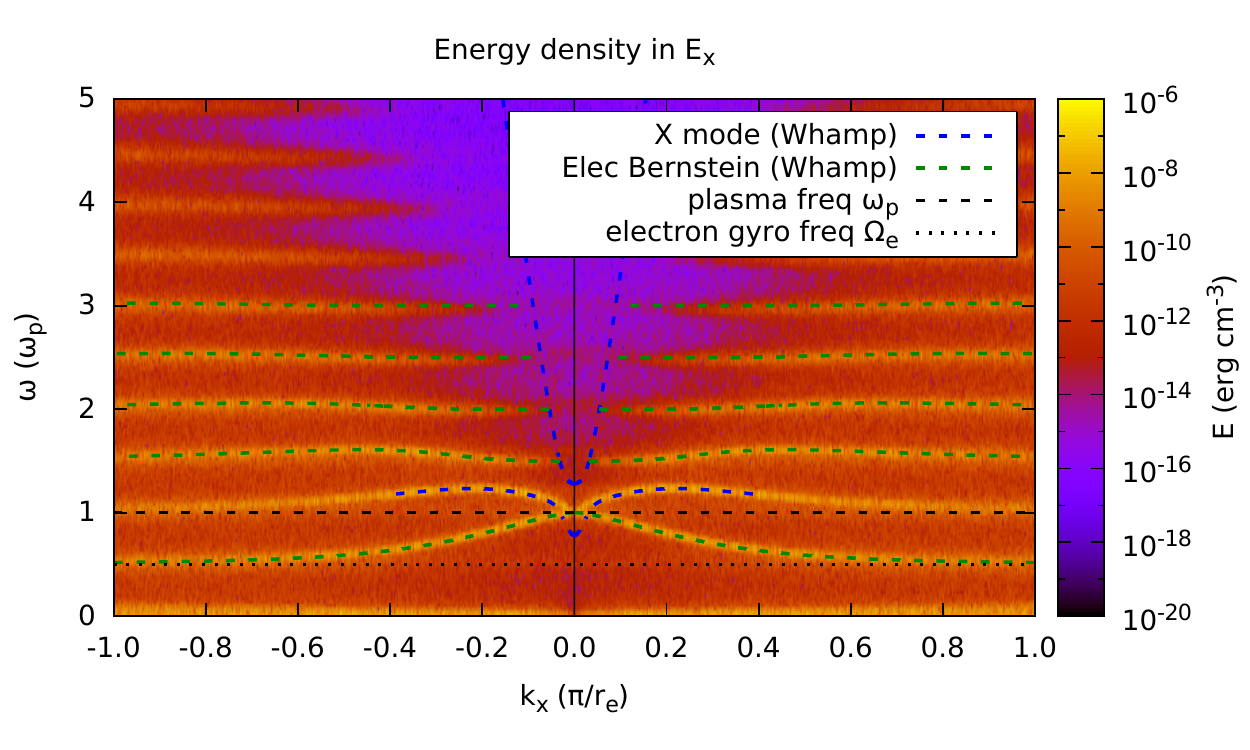}
\end{center}
\caption{Energy density in the longitudinal component of the electric field.
Unlike Fig.~\ref{fig:em_bernstein}, the field is computed using the spectral
solver used by the radiation-free and the electrostatic plasma models.}
\label{fig:es_bernstein}
\end{figure}

Fig.~\ref{fig:es_bernstein} shows the behavior in the electrostatic model with
a static background magnetic field. Again we see a spectral hole at $k = 0$.
Compared to Fig.~\ref{fig:em_bernstein}, no remnants of the X mode
are visible here.

The absence of kinetic electrons in the hybrid models, carrying individual gyro
phases, removes electron Bernstein modes.

\FloatBarrier
\subsection{Test 6: Low Frequency R Mode}
\label{subsec:test:lf_r}

So far all simulations were as cheap as Test 1. To analyze the low frequency
regime, some more effort is required. Tab.~\ref{tab:r_lf_par} lists the changes
to the simulation parameters.

\begin{table}[hbpt]
\begin{center}
\begin{tabular}{l l r}
simulation domain & $L_\mathrm{z}$ & $8192 \, \Delta x$ \\
simulation duration & $T_\mathrm{sim}$ & $1000 \, \omega_\mathrm{p,e}^{-1}$ \\
& & $52000 \, \Delta t$\\
\end{tabular}
\end{center}
\caption{Simulation size to study the low frequency R mode.}
\label{tab:r_lf_par}
\end{table}

\begin{figure}[htbp]
\begin{center}
	\includegraphics[width=\columnwidth]{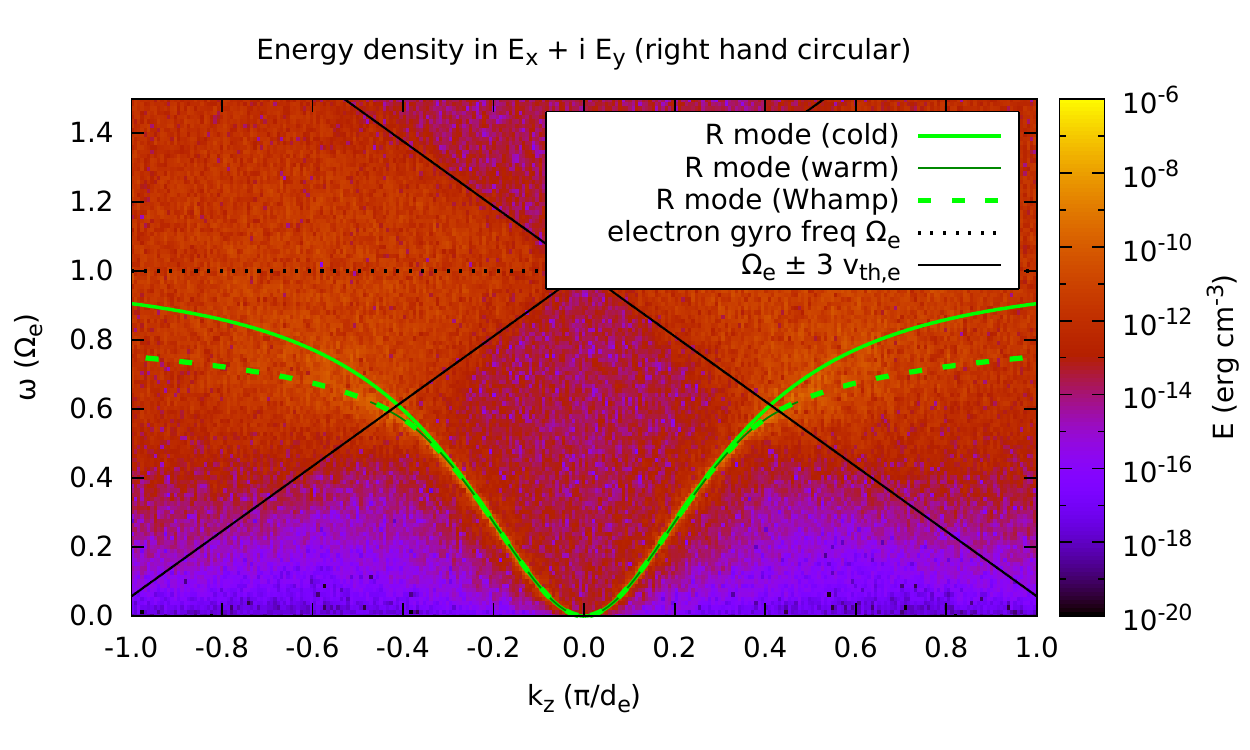}
\end{center}
\caption{Energy density in the right handed circularly polarized part of the
electric field. Similar to Fig.~\ref{fig:disp_em_hf_r}, the electric field is
computed from the electromagnetic plasma model, but this time with the
parameters given in Tabs.~\ref{tab:baseline_def} and \ref{tab:r_lf_par} to reach
the low frequency regime. As expected, the low frequency branch of the R mode is
visible, including the frequency range of whistler waves.}
\label{fig:em_lf_r}
\end{figure}

When we run the simulation using those parameters and plot the result, we again
find the L and R mode. Limiting ourselves to frequencies up to the electron
gyro frequency and filtering for right hand circular polarization, we get
Fig.~\ref{fig:em_lf_r}.

Fig.~\ref{fig:em_lf_r} is dominated by the low frequency branch of the R mode
which contains a region where $\omega$ depends quadratically on $k$. These
waves would usually be classified as whistler waves.

For larger $k$, the group velocity drops again as the wave comes closer to the
resonance at $\Omega_\mathrm{c,e}$. In this region, a deviation can be seen
between the prediction for a cold plasma and the mode in the simulated plasma
of finite temperature.  Using either the prediction of \cite{Chen_2013} for a
warm plasma or the linearized equations of WHAMP, this effect can be predicted
correctly. Both approaches, however, require to numerically solve for the
dispersion relation.

For even larger $k$, the mode is damped away by gyrating particles, which is
also predicted for warm plasmas. The gyrating electrons generate noise cones
around the electron gyro frequency that are clearly visible. The opening angle
of the cones corresponds to roughly $3\sqrt{2}~v_\mathrm{th,e}~k$.

\begin{figure}[htbp]
\begin{center}
	\includegraphics[width=\columnwidth]{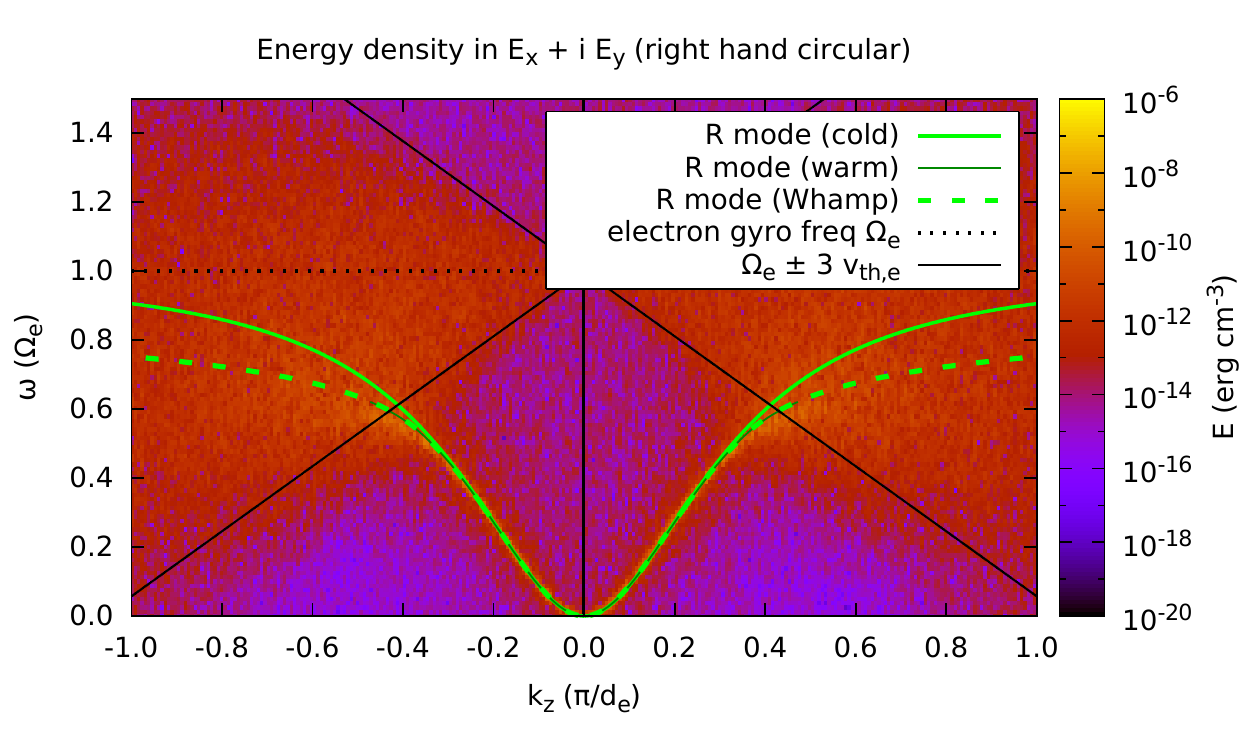}
\end{center}
\caption{Energy density in the right handed circularly polarized part of the
electric field. Unlike Fig.~\ref{fig:em_lf_r} the field is computed using the
radiation-free plasma model, but otherwise using the same parameters.}
\label{fig:darwin_lf_r}
\end{figure}

Fig.~\ref{fig:darwin_lf_r} shows the low frequency modes that are right
handed circularly polarized from a radiation free plasma model. Unlike the high
frequency branches of L and R mode that are removed when the electromagnetic
radiation is removed, the low frequency branches still exist and show the same
features as in an full electromagnetic plasma model.

In an electrostatic plasma model, these waves are missing because no transverse
fields exist.

\begin{figure}[htbp]
\begin{center}
	\includegraphics[width=\columnwidth]{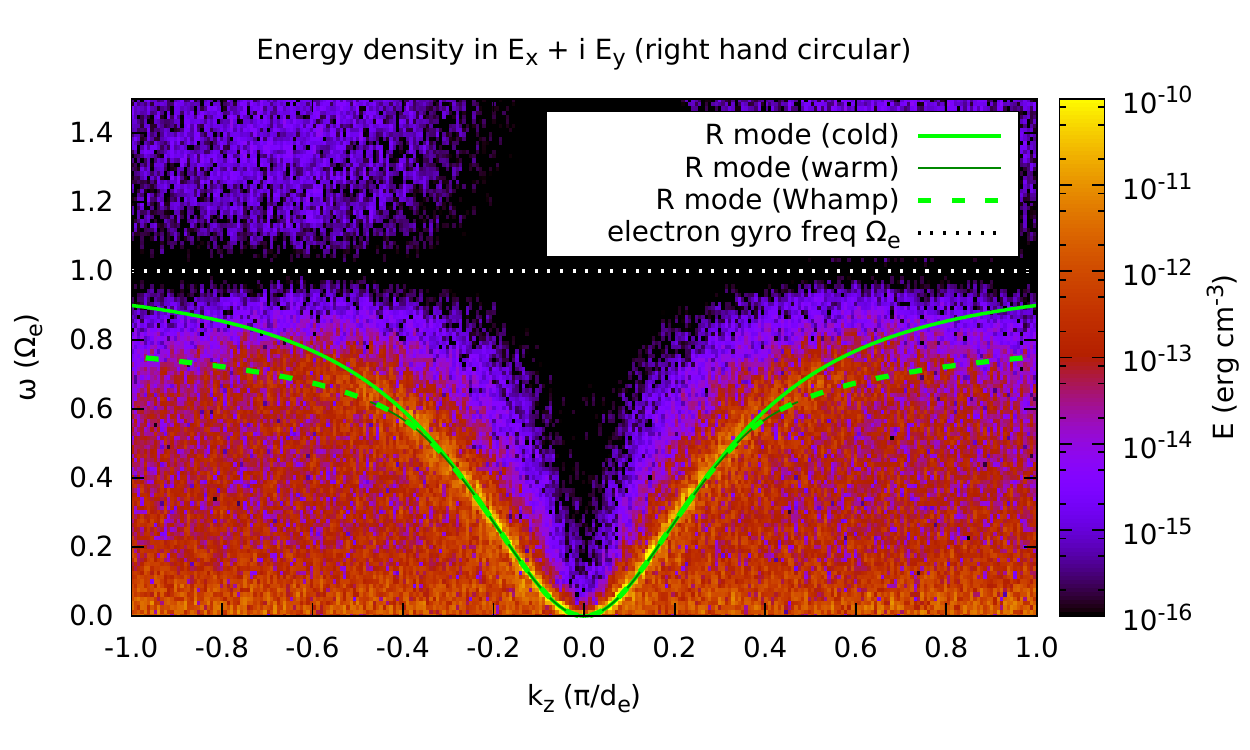}
\end{center}
\caption{Energy density in the right handed circularly polarized part of the
electric field. In this figure the field is computed from the hybrid model with
electron inertia. Compared to Fig.~\ref{fig:em_lf_r}, the noise cones around
the gyro frequency of electrons are missing.}
\label{fig:massive_lf_r}
\end{figure}

As Fig.~\ref{fig:massive_lf_r} shows, the noise cones of gyrating electrons --
a purely kinetic effect -- are missing in a model that uses an electron fluid.
The low frequency branch of the R mode, however, still exists and shows the
correct dispersion relation. This is not surprising as the wave is carried by
electrons but can be derived in the (fluid-like) plasma theory. The
gyro frequency of electrons can be estimated from the spectral gap in the noise.

\begin{figure}[htbp]
\begin{center}
	\includegraphics[width=\columnwidth]{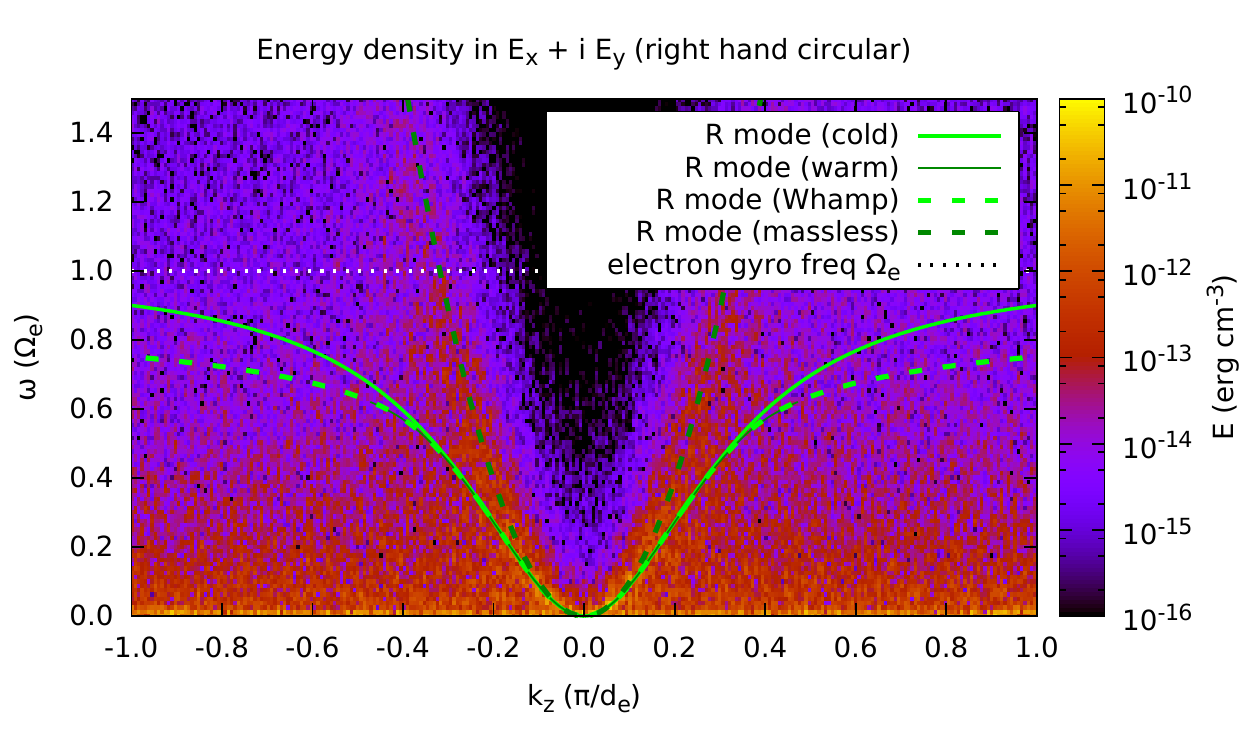}
\end{center}
\caption{Energy density in the right handed circularly polarized part of the
electric field. This time a hybrid model with massless electrons is used to
compute the electric field. The dispersion relation of the low frequency R mode
is significantly modified by this model as a comparison with
Figs.~\ref{fig:em_lf_r} to \ref{fig:massive_lf_r} shows.}
\label{fig:massless_lf_r}
\end{figure}

Fig.~\ref{fig:massless_lf_r} shows the right hand circular mode in the limit
$m_\mathrm{e} \rightarrow 0$. For small $k$, the mode is unchanged by the lack
of electron inertia, but at larger $k$ it lacks the resonance at the electron
gyro frequency, which is not well defined without electron mass.

Normalizing in the same way as used for the plot ${\omega = \tilde{\omega}\,
\Omega_\mathrm{c,e}}$ and ${k = \tilde{k} \, \pi / d_\mathrm{e}}$
simplifies the dispersion relation given in Eq.~\ref{eqn:disp_r_emhd_massless}
significantly:
\begin{equation}
	\tilde{\omega} = \pi^2 \, \tilde{k}^2 \quad ,
\label{eqn:disp_r_emhd_massless_norm}
\end{equation}
The plot of the spectral energy density indeed shows that the
wave mode follows this dispersion relation even for frequencies $\omega$ larger
than the electron gyro frequency.

\FloatBarrier
\subsection{Test 7: Ion Bernstein Modes}
\label{subsec:test:ion_bernstein}

\begin{table}[hbpt]
\begin{center}
\begin{tabular}{l l r}
electron plasma frequency & $\omega_\mathrm{p,e}$ & $1.000 \cdot 10^9\, \mathrm{rad} / \mathrm{s}$ \\
electron gyro frequency & $\Omega_\mathrm{c,e}$ & $2.144 \cdot 10^8\, \mathrm{rad} / \mathrm{s}$ \\
electron thermal speed & $v_\mathrm{th,e}$ & 0.021 c \\
mass ratio & $m_\mathrm{i} / m_\mathrm{e}$ & 1836 \\
temperature & $T$ & 2.71 MK\\
Debye length & $\lambda_\mathrm{D}$ & 0.641 cm \\
grid size & $\Delta x$ & 0.454 cm\\
electron gyro radius & $r_\mathrm{e}$ & 2.993 cm \\
magnetic field & $B_0$ & 1.219 mT\\
               &       & 12.19 G\\
Alfv\'en speed & $v_\mathrm{A}$ & 0.005 c \\
simulation domain & $L_x$ & $16384 \, \Delta x$ \\
simulation duration & $T_\mathrm{sim}$ & $150 \, \Omega_{c,i}^{-1}$
\end{tabular}
\end{center}
\caption{Simulation parameters used for the ion Bernstein modes.}
\label{tab:ion_bernstein}
\end{table}

\begin{figure}[htbp]
\begin{center}
	\includegraphics[width=\columnwidth]{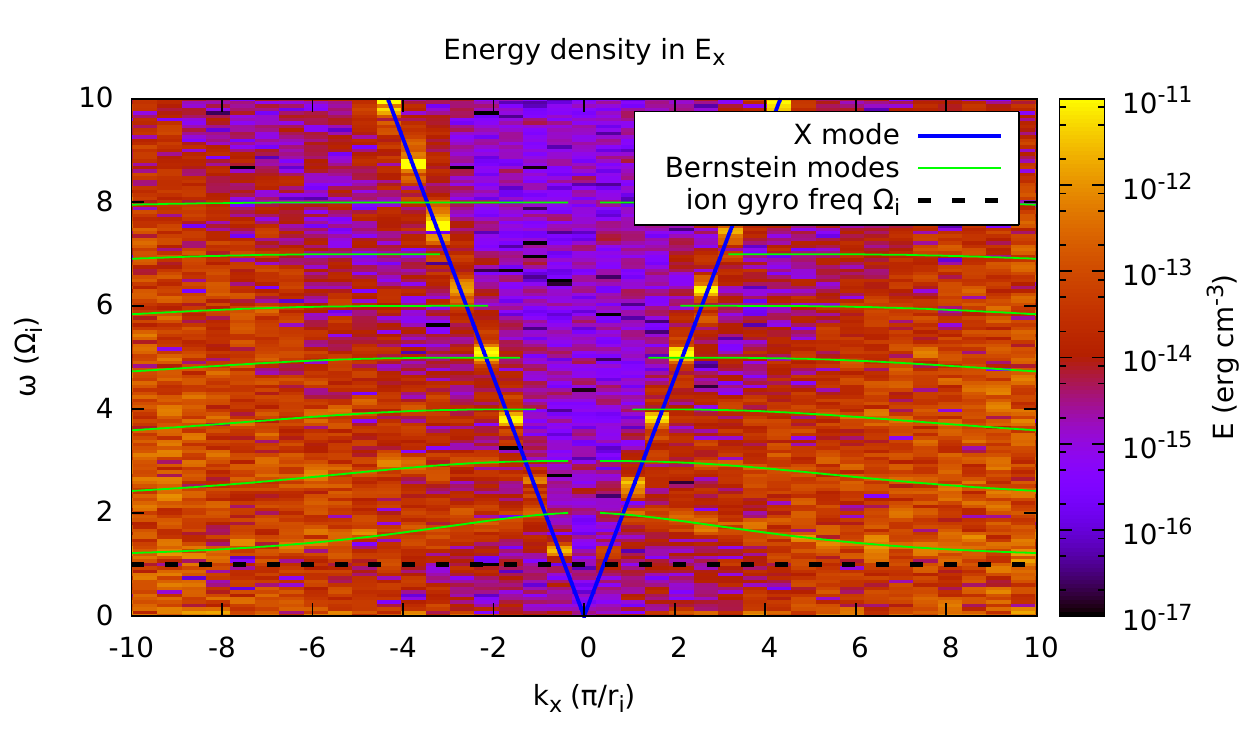}
\end{center}
\caption{Energy density in the longitudinal component of the electric field.
The field has been obtained from the electromagnetic plasma model using the
parameter set given in Tab.~\ref{tab:ion_bernstein}.}
\label{fig:em_ion_bernstein}
\end{figure}

Fig.~\ref{fig:em_ion_bernstein} shows the result of test 7 using the
electromagnetic plasma model. The simulation is computationally expensive and
quite noisy. The X mode is clearly visible. At smaller phase speeds, ion
Bernstein modes are visible. Better resolution and lower noise levels (e.g.
through a larger number of particles per cell) would be required to make this
an efficient test, but would increase the computational cost even further.

\begin{figure}[htbp]
\begin{center}
	\includegraphics[width=\columnwidth]{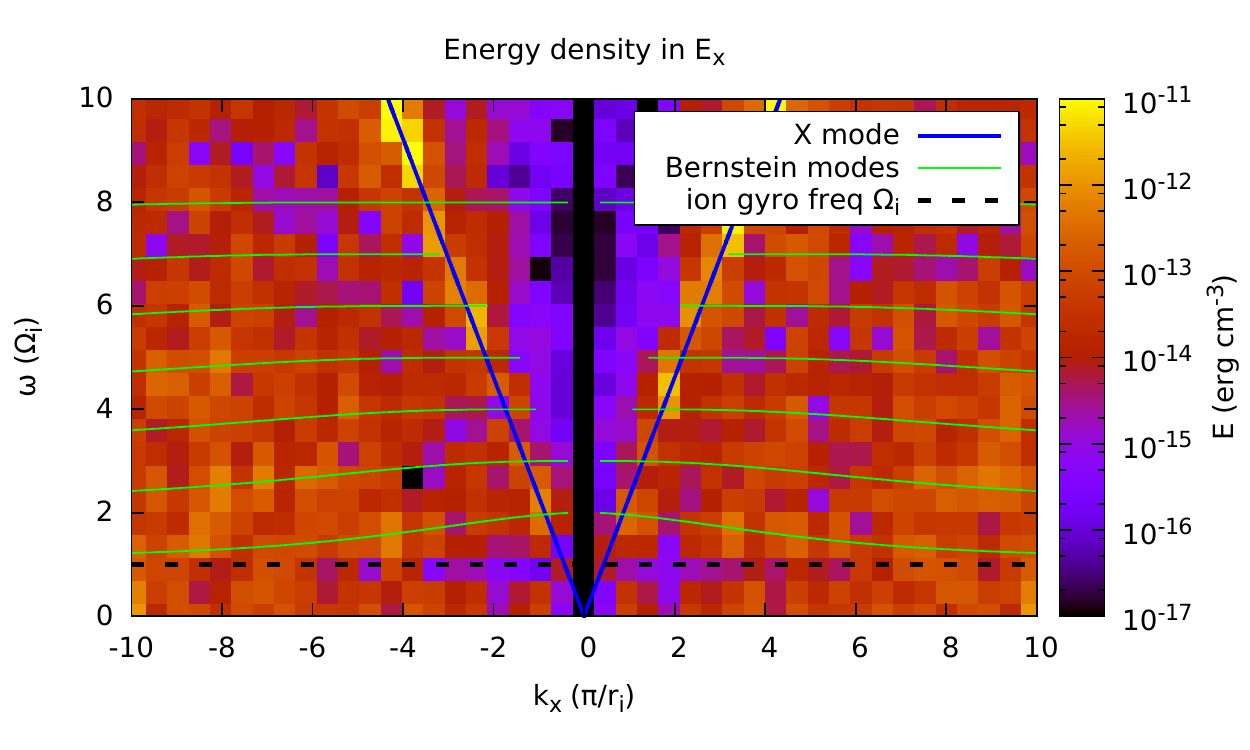}
\end{center}
\caption{Energy density in the longitudinal component of the electric field.
The field has been obtained from the spectral solver used for the
radiation-free plasma model.}
\label{fig:darwin_ion_bernstein}
\end{figure}

Fig.~\ref{fig:darwin_ion_bernstein} shows the result of the spectral solver as
used in the radiation-free plasma model. As usual with this solver, a hole in
the spectral energy density appears at $k = 0$. This plasma model allows
larger time steps than the electromagnetic model, but the simulation is still too
expensive to make an efficient test problem.

\begin{figure}[htbp]
\begin{center}
	\includegraphics[width=\columnwidth]{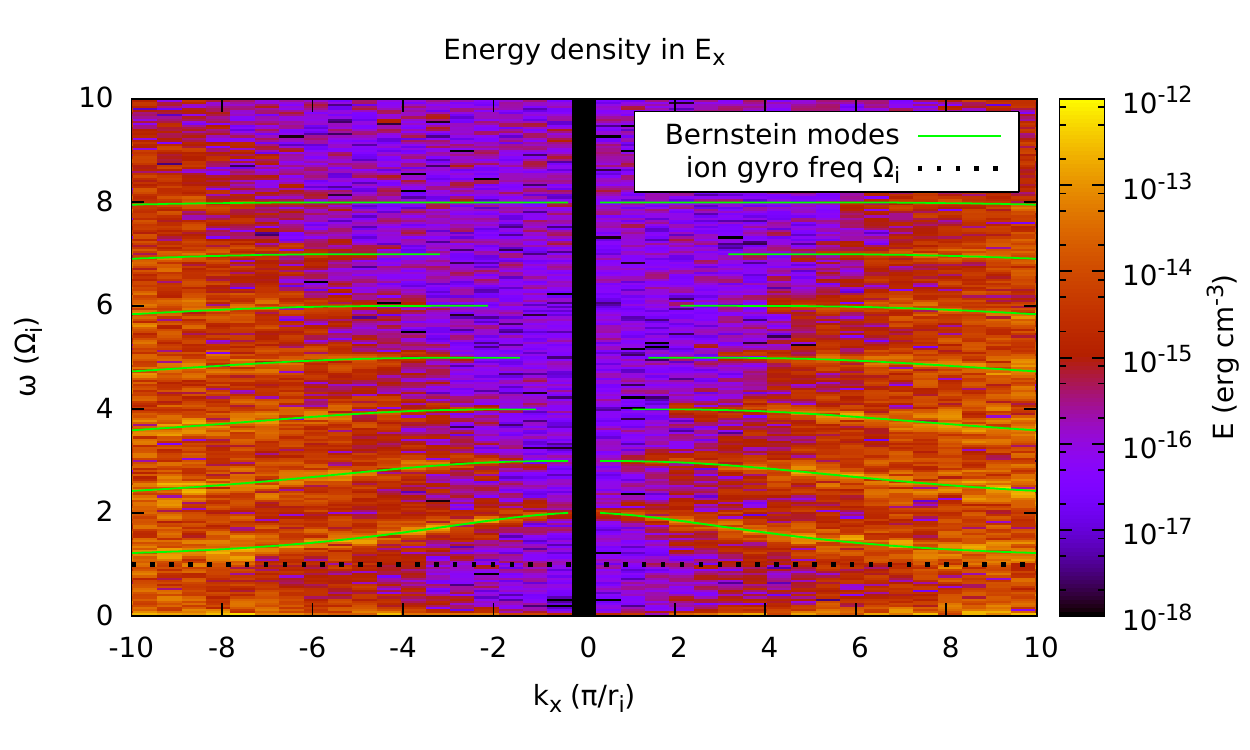}
\end{center}
\caption{Energy density in the longitudinal component of the electric field.
The field has been obtained from the spectral solver used for the electrostatic
plasma model using the parameter set given in Tab.~\ref{tab:ion_bernstein}. The
only visible wave modes are ion Bernstein waves.}
\label{fig:es_ion_bernstein}
\end{figure}

Fig.~\ref{fig:es_ion_bernstein} shows the result of the spectral solver as used
in the electrostatic plasma model. This model does not include the X mode and
allows much larger time steps, which reduces computational expense. Additionally,
a single time is cheaper than in the radiation-free model and the
test does not rely on the transverse field components that are missing in the
electrostatic model.

\begin{figure}[htbp]
\begin{center}
	\includegraphics[width=\columnwidth]{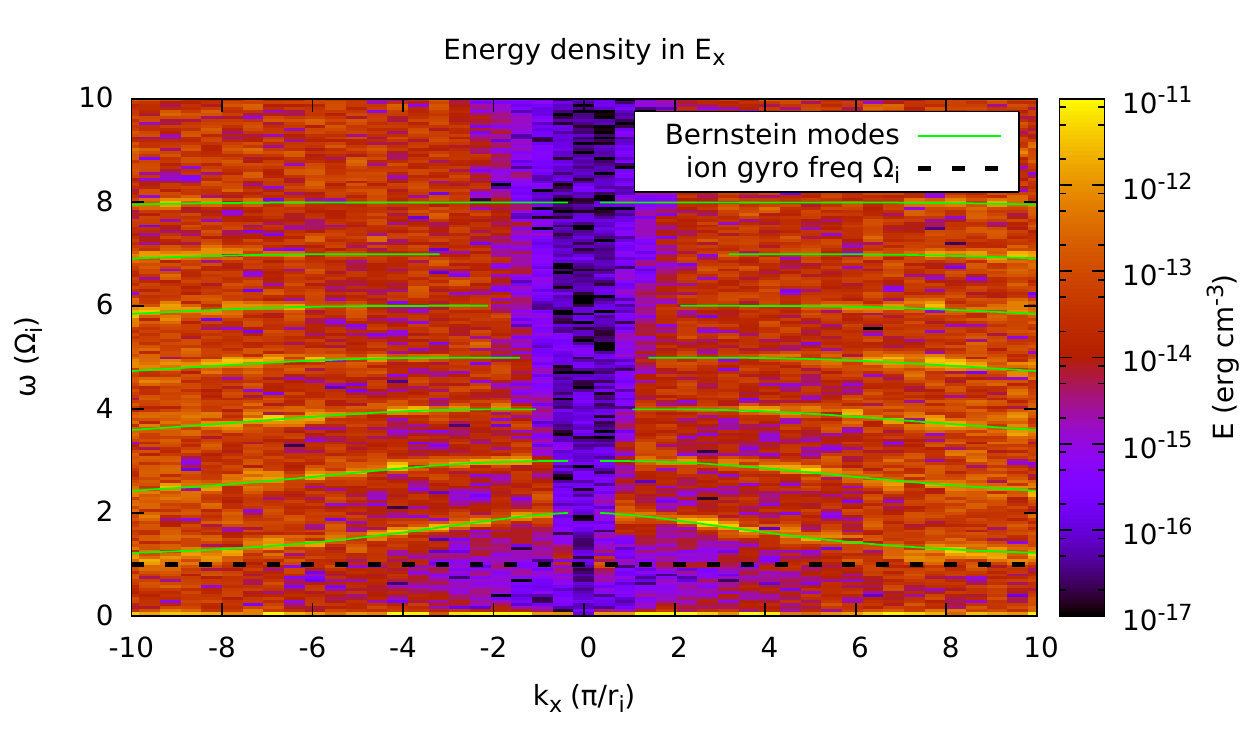}
\end{center}
\caption{Energy density in the longitudinal component of the electric field.
The plot is based on the hybrid model with electron inertia and the parameter
set given in Tab.~\ref{tab:ion_bernstein}.}
\label{fig:massive_ion_bernstein}
\end{figure}

Fig.~\ref{fig:massive_ion_bernstein} shows the output of the hybrid plasma
model including effects of electron inertia. In this model, ions are treated as
kinetic particles and, as expected, ion Bernstein modes are visible. Note the
reappearance of the X mode as an enhanced band of noise at relatively large
phase velocities.

\begin{figure}[htbp]
\begin{center}
	\includegraphics[width=\columnwidth]{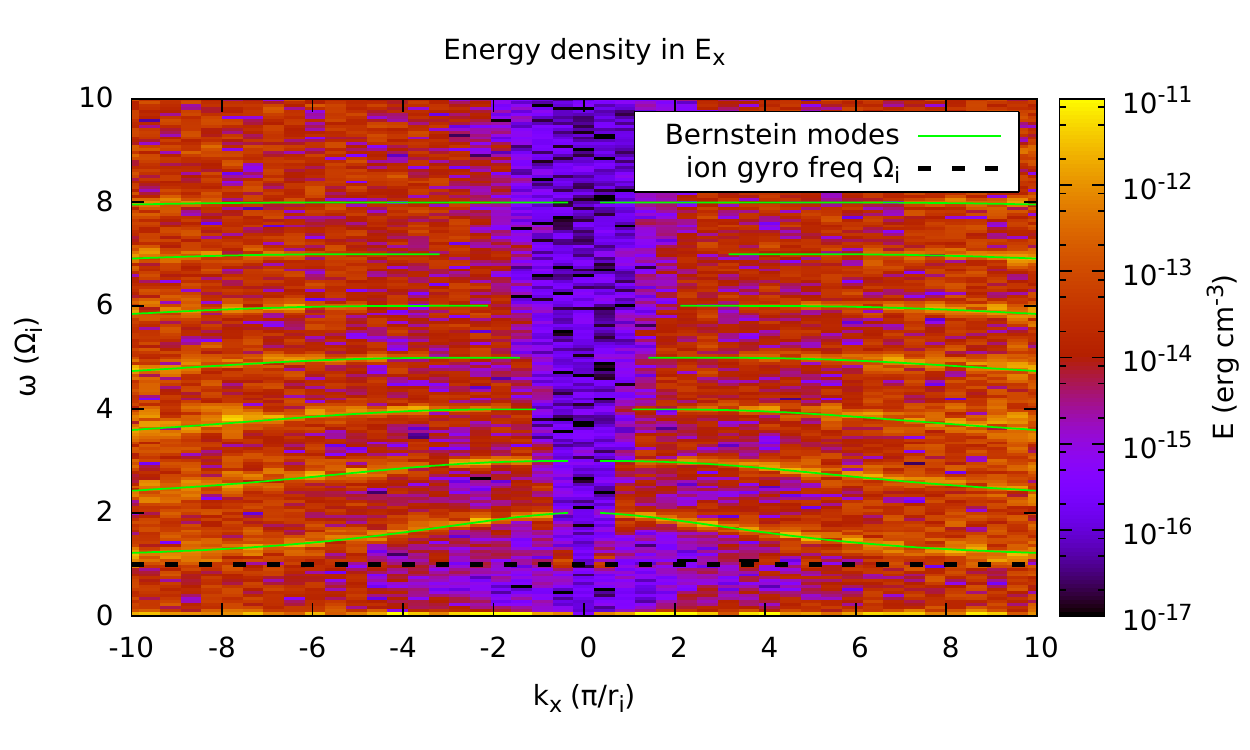}
\end{center}
\caption{Energy density in the longitudinal component of the electric field.
The hybrid model used the parameter set given in Tab.~\ref{tab:ion_bernstein}
and massless electrons. The case including electron inertia can be found in
Fig.~\ref{fig:massive_ion_bernstein}.}
\label{fig:massless_ion_bernstein}
\end{figure}

Fig.~\ref{fig:massless_ion_bernstein} shows the hybrid plasma models without
electron inertia. The ion Bernstein modes are dominated by ion kinetic effects
and remain unchanged. Given that this plasma model admits a only limited number
of modes, this is probably the most relevant test problem for it.

\FloatBarrier
\subsection{Test 8: Low Frequency L Mode}
\label{subsec:test:lf_l}

Resolving low frequency L modes, as done in Sec.~\ref{subsec:test:lf_r} for
their right handed counterparts, would require another large increase in
effort.  (One needs 1836 times as many time steps to resolve the lower gyro
frequency and $\sqrt{1836}$ more cells to capture the larger gyro radius.) The
only feasible way is to reduce the mass ratio between protons and electrons.
Low mass ratios result in possibly unrealistic high Alfv\'en speeds if the
magnetic field is not adjusted. Tab.~\ref{tab:l_lf_par} shows parameters that
are a reasonable tradeoff and allow a glimpse at this wave mode.

\begin{table}[hbpt]
\begin{center}
\begin{tabular}{l l r}
mass ratio & $m_\mathrm{i} / m_\mathrm{e}$ & 18.36 \\
Alfv\'en speed & $v_\mathrm{A}$ & 0.117 c\\
simulation domain & $L_\mathrm{z}$ & $16384 \, \Delta x$ \\
simulation duration & $T_\mathrm{sim}$ & $4000 \, \omega_\mathrm{p,e}^{-1}$ \\
& & $208000 \, \Delta t$\\
particle updates & & $3.5 \cdot 10^{12}$\\
\end{tabular}
\end{center}
\caption{Simulation size to study the low frequency L mode.}
\label{tab:l_lf_par}
\end{table}

\begin{figure}[htbp]
\begin{center}
	\includegraphics[width=\columnwidth]{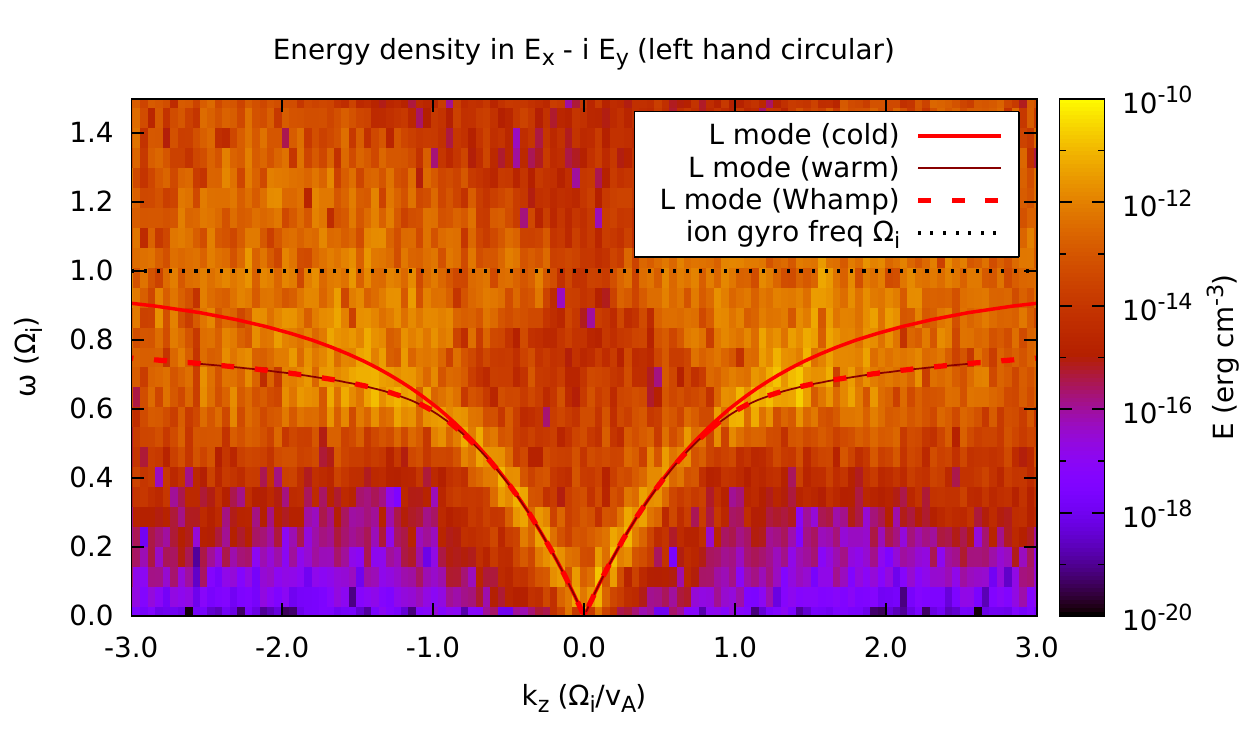}
\end{center}
\caption{Energy density in the left handed circularly polarized part of the
electric field. The plot is based on a simulation using the electromagnetic
plasma model and parameters given in Tab.~\ref{tab:l_lf_par}. As expected, the
low frequency L mode is visible as well as noise cones produced by gyrating
ions.}
\label{fig:em_lf_l}
\end{figure}

Running the simulation with those parameters and plotting the spectral energy
density of left handed modes results in Fig.~\ref{fig:em_lf_l}. The low
frequency branch of the L mode is clearly visible. For small $k$, it matches
well the prediction for a cold plasma. For intermediate $k$, effects of the
finite temperature have to be included to explain the simulation results. At
higher $k$, the mode is damped away by gyrating protons that are visible as
noise cones. These cones are analogous to the cones generated by gyrating
electrons, but occur on ion scales, i.e. they are centered on the gyro
frequency of the ions and the opening is determined by the thermal speed of
ions.

\begin{figure}[htbp]
\begin{center}
	\includegraphics[width=\columnwidth]{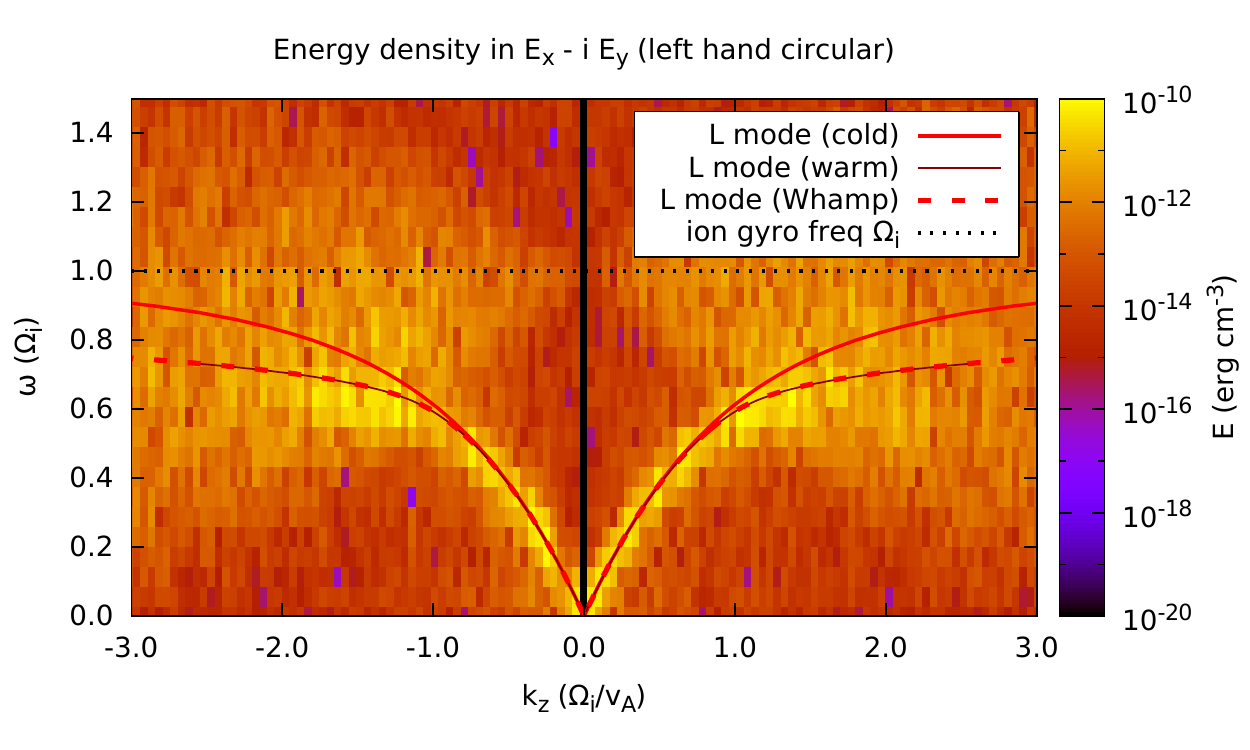}
\end{center}
\caption{Energy density in the left handed circularly polarized part of the
electric field obtained from the radiation-free plasma model. The simulation
parameters can be found in Tab.~\ref{tab:l_lf_par}.}
\label{fig:darwin_lf_l}
\end{figure}

In Fig.~\ref{fig:darwin_lf_l} it can be seen that the left handed low frequency
waves survive in the radiation-free plasma, the same as the right handed
counterparts.

\begin{figure}[htbp]
\begin{center}
	\includegraphics[width=\columnwidth]{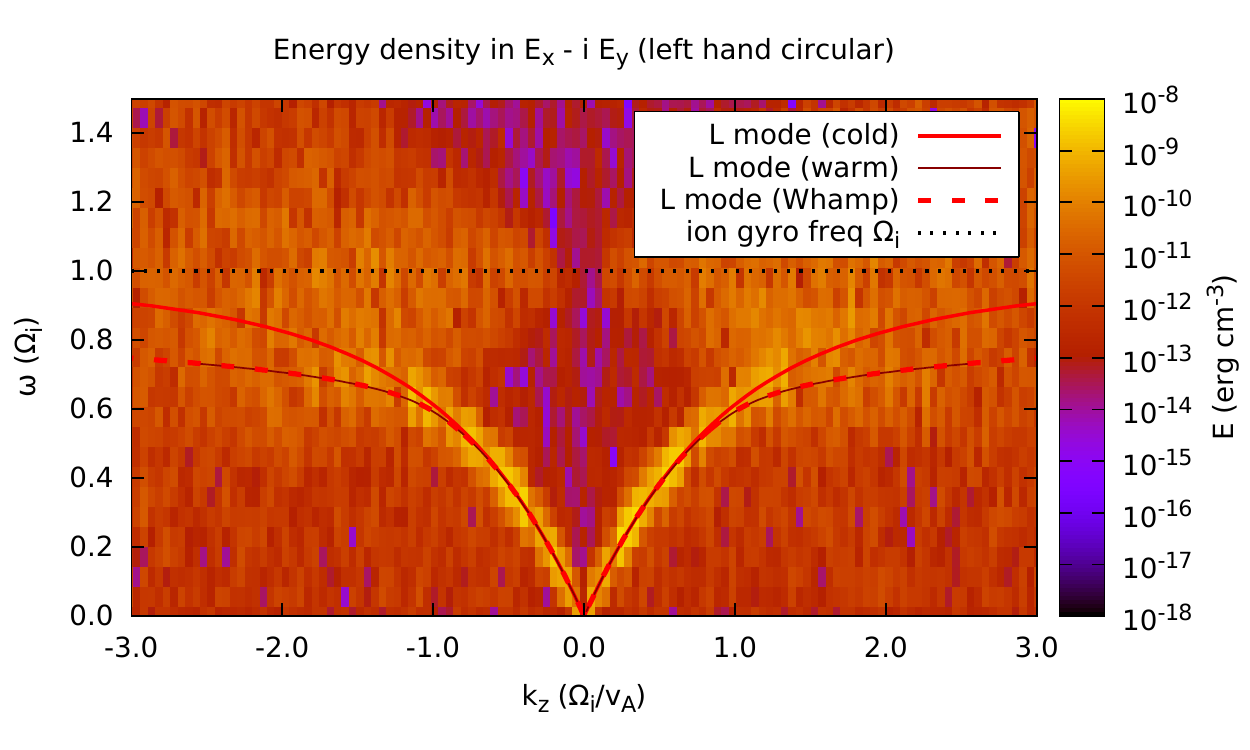}
\end{center}
\caption{Energy density in the left handed circularly polarized part of the
electric field. Again parameters from Tab.~\ref{tab:l_lf_par}, but this time
for a hybrid model with electron inertia. The ions are still treated
kinetically and consequently the low frequency L mode and the noise cones are
retained.}
\label{fig:massive_lf_l}
\end{figure}

\begin{figure}[htbp]
\begin{center}
	\includegraphics[width=\columnwidth]{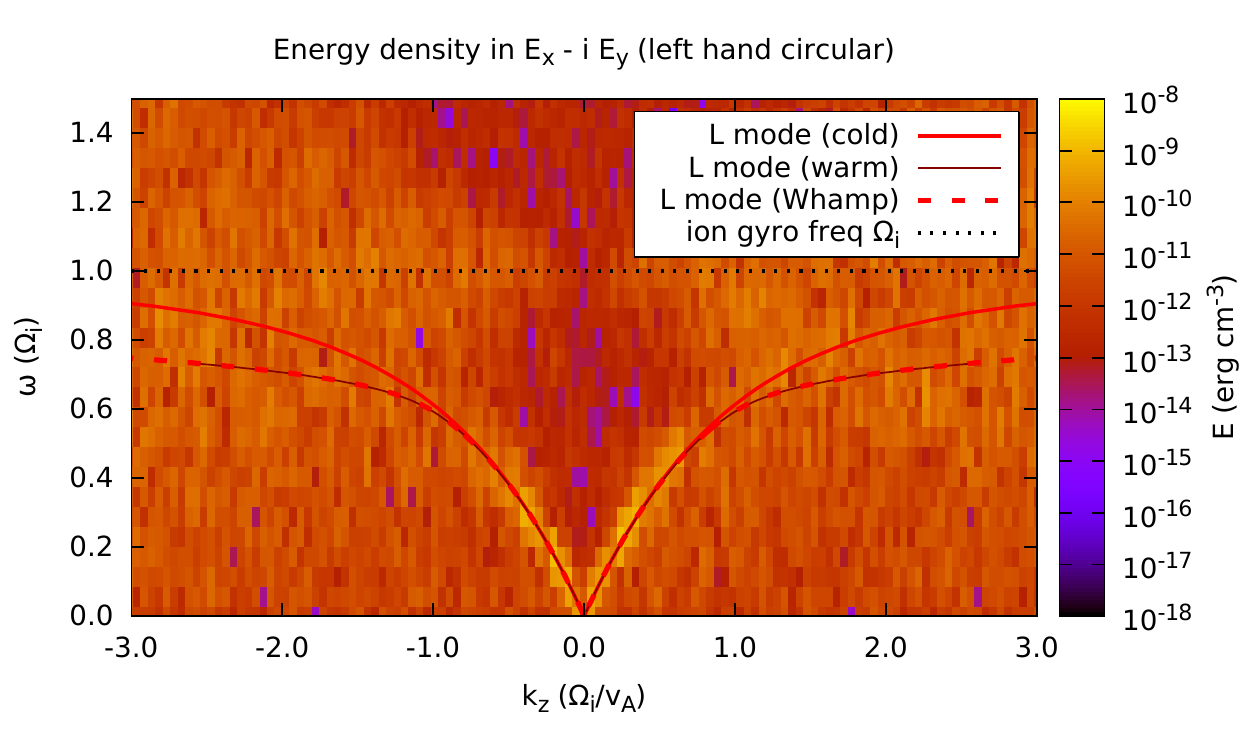}
\end{center}
\caption{Energy density in the left handed circularly polarized part of the
electric field. Compared to Fig.~\ref{fig:massive_lf_l} electron inertia has
been removed.}
\label{fig:massless_lf_l}
\end{figure}

Figs.~\ref{fig:massive_lf_l} and \ref{fig:massless_lf_l} show that the low
frequency L mode branches exist basically unaltered without kinetic electrons.
The noise cones of gyrating ions are unaffected unlike
Fig.~\ref{fig:massive_lf_r}.

\FloatBarrier
\section{Conclusions}
\label{sec:conclusion}

In this work we proposed a set of test problems suitable for a wide range of
kinetic plasma models and provided reference results based on our numerical
codes. Those tests are based on a set of different plasma wave modes and are
useful to check quickly and conveniently the correct implementation of
different plasma models, including the correct interaction of the different
parts of the simulation program handling particles and electromagnetic fields.

\begin{table*}[hbt]
\begin{center}
\begin{tabular}{l | c c c c c c c c}
Plasma Model       & \multicolumn{7}{c}{Wave Mode}\\
                   & EM & HF L/R & X & Langmuir & EB & LF R & IB & LF L\\
\hline
Electromagnetic    & X  & X      & X & X        & X  & X    &\$ &\$ \\
Radiation-free     & -  & -      & - & X        & X  & X    &\$ &\$ \\
Electrostatic      & -  & -      & - & X        & X  & -    & X & - \\
Hybrid w/ inertia  & -  & -      & - & -        & -  & X    & X & X \\
Hybrid w/o inertia & -  & -      & - & -        & -  & {\raise.17ex\hbox{$\scriptstyle\sim$}} & X & X\\
\end{tabular}
\end{center}
\caption{Suitability of modes for different plasma models. EB and IB stand for
Bernstein modes of electrons and ions, respectively. Cases indicated by 'X'
allow for an effective test. An entry of '-' indicates that the wave mode is
not suitable for testing implementations of this plasma model. The four cases
that are marked \$ are in principle suitable, but computationally expensive.
The one special case indicated with {\raise.17ex\hbox{$\scriptstyle\sim$}} is
explained in the text.}
\label{tab:test_matrix}
\end{table*}

Tab.~\ref{tab:test_matrix} shows which wave modes are suitable to test codes
implementing different plasma models.
Listed from left to right are the electromagnetic mode from
Sec.~\ref{subsec:test:electromagnetic}, left and right hand circular wave
modes at or above the plasma frequency (Sec.~\ref{subsec:test:hf_lr}), the
extraordinary mode (Sec.~\ref{subsec:test:extraordinary}), the Langmuir mode
(Sec.~\ref{subsec:test:langmuir}), electron Bernstein modes
(Sec.~\ref{subsec:test:electron_bernstein}), low frequency waves with right
hand circular polarization (Sec.~\ref{subsec:test:lf_r}),  ion Bernstein
modes (Sec.~\ref{subsec:test:ion_bernstein}) and low frequency left hand
circular polarization (Sec.~\ref{subsec:test:lf_l}).
Listed on the left hand side are the different plasma models -- from top to
bottom --, the electromagnetic model (Sec.~\ref{subsec:electromagnetic}),
the radiation-free model (Sec.~\ref{subsec:darwin}), the electrostatic model
(Sec.~\ref{subsec:electrostatic}) and the model using an implicit electron
fluid -- either with or without electron inertia -- described in
Sec.~\ref{subsec:hybrid}.

Wave modes that provide suitable test problems for the chosen plasma model are
indicated with 'X'. If a '-' is listed, the wave mode is not present or usable
in the plasma model.
For two plasma models, the low frequency left hand circularly polarized waves
and the ion Bernstein modes are marked with \$. These waves do exist in the
electromagnetic and radiation-free plasma model and show the properties
expected from cold plasma theory. In principle, these waves could be used to
test the simulation code, e.g. by extracting the gyro frequency of ions or the
Alfv\'en velocity.  Simulations with sufficient resolution are, however,
computationally very expensive. Given that these plasma models admit a large
number of alternative wave modes, it is better to choose an alternative test
problem unless low frequency properties of the ions are explicitly needed.
A special case (indicated by '{\raise.17ex\hbox{$\scriptstyle\sim$}}'), occurs
for low frequency right hand circularly polarized waves in the hybrid model
without electron inertia. As shown in Fig.~\ref{fig:massless_lf_r}, such a wave
mode does exist, however, the dispersion relation is modified compared to all
other plasma models used here. In particular the resonance at the electron gyro
frequency is missing, as it has been ordered out of the model and cannot be used
for comparison purposes. Using the modified dispersion relation in
Eq.~\ref{eqn:disp_r_emhd_massless}, it is possible to recover the combination
of magnetic field and electron density. Given the limited number of suitable
wave modes in this hybrid plasma model, right hand circularly polarized waves
are still an important test problem, but analysis requires extra attention, and
direct comparison with other plasma models is difficult.

\section*{Acknowledgment}

We acknowledge the use of the ACRONYM code and would like to thank the
developers (Verein zur F\"orderung kinetischer Plasmasimulationen e.V.) for
their support. The implementation of the hybrid model is based on a combination
with the EMHD code of Neeraj Jain. The authors would like to thank him for all
the work on the joint code and all the useful discussion on hybrid plasma
models.

P.M. acknowledges financial support by the Max-Planck-Princeton Center for
Plasma Physics. This work is based upon research supported by the National
Research Foundation and Department of Science and Technology. Any opinion,
findings and conclusions or recommendations expressed in this material are
those of the authors and therefore the NRF and DST do not accept any liability
in regard thereto.


\end{document}